# QUANTUM GEOMETRY AND GRAVITY*

EDUARD PRUGOVEČKI
*Department of Mathematics*
*University of Toronto*
*Toronto, Canada M5S 1A1*

**ABSTRACT**

The geometro-stochastic method of quantization provides a framework for quantum general relativity, in which the principal frame bundles of local Lorentz frames that underlie the fibre-theoretical approach to classical general relativity are replaced by Poincaré-covariant quantum frame bundles. In the semiclassical regime for quantum field theory in curved spacetime, where the gravitational field is not quantized, the elements of these local quantum frames are generalized coherent states, which emerge naturally from phase space representations of the Poincaré group. Due to their informational completeness, these quantum frames are capable of taking over the role played by complete sets of observables in conventional quantum theory. The propagation of quantum-geometric fields proceeds by path integral methods, based on parallel transport along broken paths consisting of arcs of geodesics of the Levi-Civita connection. The formulation of quantum gravity within this framework necessitates the transition to quantum superframe bundles and a quantum gravitational supergroup capable of incorporating diffeomorphism invariance into the framework. This results in a geometric version of quantum gravity which shares some conceptual features with covariant as well as with canonical gravity, but which avoids the foundational and the mathematical difficulties encountered by these two approaches.

## 1. Introduction

During the preceding decade a new method of quantization was developed[1-4] in order to deal with some of the unresolved foundational problems[5,6] in relativistic quantum particle and field localization. This geometro-stochastic method eventually led[7,8] to a purely geometric framework for quantum general relativity, and proved capable of resolving foundational problems in quantum field theory in curved spacetime and in quantum gravity .[9,10]

The geometro-stochastic method of quantization shares[11] some features with the well-known geometric method of Kostant[12] and Souriau[13] in that it employs phase space representations of the Galilei and the Poincaré group, albeit of a different type[14] than the Kostant and Souriau method. In fact, the geometro-stochastic method is intrisically based on coherent states, and in this respect it displays common features (cf. Ref. 15, Sec. 16.2) with the method of Berezin.[16] However, in its scope,[9,10] it transcends both these methods, since it is applicable to quantum field theory (QFT) in the general relativistic context.

This is due to the fact that the geometro-stochastic quantization is based on a concept of quantum frame capable of taking over in the general relativistic regime, physically as well as mathematically, the role played in conventional quantum theory by complete sets of observables. In QFT in a given classical curved spacetime, principal bundles of such Poincaré quantum frame mediate the formulation of quantum geometries that are Hilbert or pseudo-Hilbert fibre bundles associated with principal frame bundles over curved spacetime. In turn, this enables a purely geometric type of formulation of propagation of

---

quantum fields in such bundles by means of path integrals that conform to the strong equivalence principle of general relativity.[17] The corner stone on which this kind of propagation is based consists of quantum connections whose connection coefficients are provided by the Levi-Civita connection form in the given curved spacetime. Locally, such quantum geometric propagation is governed by energy density operators which arise from well-defined stress-energy tensor operators, which in the special relativistic regime give rise to infinitesimal generators of spacetime translations and Lorentz transformations.

The relation of this quantum-geometric framework for QFT to the conventional framework is arrived at by establishing from purely geometric considerations the existence of action integrals. A consistent physical interpretation of this framework is then obtained by proving the existence of Poincaré gauge invariant and locally conserved probability currents.

The extrapolation of these physical ideas and mathematical techniques to a geometric formulation of quantum gravity requires the use of geometric versions of BRST symmetries in superfibres over a quantum spacetime supermanifold.[9,10]

In Sec. 2 we describe those basic features of the geometro-stochastic method of quantization in the nonrelativistic regime which demonstrate its physical consistency with orthodox quantum mechanics. We then describe in Sec. 3 the special relativistic counterparts of these features. In Sec. 4 we extrapolate this framework into a geometric framework for quantum field theory in curved spacetime. In Sec. 5 we describe the basic features of the quantum-geometric mode of field propagation formulated within such a framework. We then outline in Sec. 6, in very general terms, how these physical ideas and techniques can be applied to quantum gravity; whereas, in the remaining Secs. 7-11 we highlight and explain the principal mathematical aspects of this application.

## 2. Nonrelativistic Geometro-Stochastic Quantization

It is well-known that the canonical quantization of a classical Hamiltonian system carried out in general canonical coordinate leads to results that are generically inconsistent with those obtained in the Cartesian coordinates related the adopted class of inertial reference frames. Therefore, the fundamental thesis underlying the geometro-stochastic method of quantization is that it is *the choice of reference frames that is of primary significance to the quantization procedure*, and not that of coordinates in general; moreover, that the choice of coordinates can be signicant only if they are operationally defined in relation to those frames. This suggests basing quantization on a group-theoretical foundation, in which the *operational* procedures required during the construction of *inertial* frames receives a natural interpretation that is directly related to the group of physical operations describing changes of such frames, and to the operational procedures carried out with them.

These fundamental physical ideas lead to the geometro-stochastic method of quantization, which, for the sake of simplicity, we shall describe for the spin-0 case. In the nonrelativistic regime the Galilei group plays the role of fundamental kinematical group. Moreover, the elements $(\mathbf{q},\mathbf{p})$ of the nonrelativistic phase space $\Gamma$ are represented by the Cartesian coordinate triples $\mathbf{q},\mathbf{p}\in\mathbf{R}^3$ in relation to any given classical inertial frame of reference $\boldsymbol{u}$. The geometro-stochastic quantization scheme first introduced in Appendix C of Ref. 3 requires that representations of the canonical commutation relations with respect to $\boldsymbol{u}$ are to be derived from the irreducible subrepresentation of the unitary ray representation of the Galilei group in $L^2(\Gamma)$. Hence, let us denote by $(b,\mathbf{a},\mathbf{v},R)$ the generic element of that group, with $b$ being the amount a time translation, the 3-vectors $\mathbf{a}$ and $\mathbf{v}$ representing a space translation and a velocity boost, respectively, and $R$ a being a 3-rotation. Then the required representation is obtained by assigning to $(b,\mathbf{a},\mathbf{v},R)$ the unitary operator

$$\big(U(b,\mathbf{a},\mathbf{v},R)\,\psi\big)(\mathbf{q},\mathbf{p},t)$$
$$= \exp[i(-\tfrac{1}{2}m\mathbf{v}^2(t-b) + m\mathbf{v}\cdot(\mathbf{q}-\mathbf{a}))]\ \psi(R^{-1}[\mathbf{q}-\mathbf{a}-\mathbf{v}(t-b)],R^{-1}[\mathbf{p}-m\mathbf{v}],t-b)\ , \quad (2.1)$$

which acts on wave functions $\psi(\mathbf{q},\mathbf{p},t)$ that are, for each fixed value of $\mathbf{p}\in\mathbf{R}^3$, solutions of the free Schrödinger equation in $\mathbf{q}$ and $t$ for the mass $m$ (all expressed in Planck natural units), and which represent for each fixed value of $t$ elements of the Hilbert space $L^2(\Gamma)$ of functions which are Lebesgue square-integrable[18] in $(\mathbf{q},\mathbf{p})\in\mathbf{R}^6$.

The complete harmonic analysis of the above representation first carried out in Ref. 14 reveals an infinity of irreducible subrepresentations which are unitarily equivalent to the well-known irreducible ray representations of the Galilei group in configuration or momentum space. Each such irreducible phase space subrepresentation can be uniquely characterized by means of a unique resolution generator $\xi\in L^2(\Gamma)$ that gives rise, in conjunction with the elements $U(b,\mathbf{a},\mathbf{v},R)$ of that representation, to the generalized coherent states

$$\xi_{\mathbf{q},\mathbf{p}} = U(0,\mathbf{q},\mathbf{p}/m,I)\xi \in L^2(\Gamma)\ , \qquad (\mathbf{q},\mathbf{p})\in\Gamma\ . \quad (2.2)$$

In turn, these generalized coherent states supply the orthogonal projector

$$\mathbf{P}_\xi = \int_{\mathbf{R}^6} \left|\xi_{\mathbf{q},\mathbf{p}}\right\rangle d\mathbf{q}d\mathbf{p}\left\langle\xi_{\mathbf{q},\mathbf{p}}\right|\ , \quad (2.3)$$

of $L^2(\Gamma)$ onto the subspace which carries that subrepresentation. The family of generalized coherent states in Eq. (2.2) is therefore uniquely assigned to the global classical frame of reference $\boldsymbol{u}$. Consequently, it can be envisaged[4,9] as constituting a quantum frame, operationally obtained by subjecting identical duplicates of a *quantum* test body, which has $\xi$ as a *proper* state vector, to the basic kinematical procedures of spacetime translation, spatial rotation and velocity boost, in order to obtain an array of kinematically correlated microdetectors that can be then used for the spatio-temporal localization of quantum systems.

This interpretation is supported by the fact that the transition amplitudes

$$\psi(\mathbf{q},\mathbf{p}) = \langle U(0,\mathbf{q},\mathbf{p}/\mathbf{m},I)\xi\,|\,\psi\rangle\ , \qquad \psi\in\mathbf{P}_\xi L^2(\Gamma)\ , \quad (2.4)$$

between the normalized quantum state vector $\psi$ of a quantum system and the constituents of such a quantum frame are directly related (cf. Ref. 9, Sec. 3.7) to the purely geometric concept of Fubini-Study distance between their representative rays in the projective space of unit Hilbert rays corresponding to elements in that subspace, and that the square of the absolute value in Eq. (2.4) can be interpreted as the probability density for detection in relation to these quantum frame elements. This interpretation follows from the fact that there is[2,3,14] a unitary map which assigns to each time-dependent wave function $\psi(\mathbf{q},\mathbf{p},t)$ in Eq. (2.1) a unique solution $\psi(\mathbf{x},t)$ of the free Schrödinger equation representing a wave function in configuration space. It can be then easily proved[4] that the probability density derivable from this phase-space wave function is related to the probability density derivable from the corresponding configuration space wave function as follows:

$$\rho_{(\xi)}(\mathbf{q},t) = \int_{\mathbf{R}^3}|\psi(\mathbf{q},\mathbf{p},t)|^2 d\mathbf{p} = (2\pi)^3\int_{\mathbf{R}^3}|\psi(\mathbf{x},t)|^2|\xi(\mathbf{x}-\mathbf{q})|^2 d\mathbf{x}\ . \quad (2.5)$$

Hence, in the *sharp-point limit*

$$(2\pi)^3|\xi(\mathbf{x}-\mathbf{q})|^2 \quad\to\quad \delta^3(\mathbf{x}-\mathbf{q})\ , \quad (2.6)$$

physically corresponding to the limiting choice of *perfectly* pointlike quantum test bodies, the probability density in Eq. (2.5) converges to the probability density for infinitely precise position measurement outcomes originally postulated by Born,

$$\rho_{(\xi)}(\mathbf{q},t) \quad \to \quad \left|\psi(\mathbf{q},t)\right|^2 \ , \tag{2.7}$$

at all those points at which this configuration space wave function is continuous.

Of special significance for the extrapolation to the special relativistic regime in Sec. 3 are the resolution generators in Eq. (2.2) which, in the configuration space representation for the Galilei group, are supplied by the ground-state wave function,

$$\xi^{(\ell)}(\mathbf{x}) = (8\pi^3\ell^2)^{-3/4}\exp(-\mathbf{x}^2/4\ell^2) \ , \qquad \ell > 0 \ , \tag{2.8}$$

of a nonrelativistic harmonic oscillator. As a matter of fact, the special relativistic counterpart of each such resolution generator $\xi^{(\ell)}$ is the ground states of a corresponding relativistic harmonic oscillator that can be interpreted[4] as the ground state of the quantum metric operator introduced by Born[5], in which $\ell$ plays the role of fundamental length. Probability currents to which such a $\xi^{(\ell)}$ gives rise,

$$\mathbf{j}_\ell(\mathbf{q},t) = \int_{\mathbf{R}^3} \frac{\mathbf{p}}{m}\left|\psi_\ell(\mathbf{q},\mathbf{p},t)\right|^2 d\mathbf{p} \ , \qquad \psi_\ell(\mathbf{q},\mathbf{p}) = \left\langle U(0,\mathbf{q},\mathbf{p}/\mathbf{m},I)\xi^{(\ell)}\middle|\psi_\ell\right\rangle \ , \tag{2.9}$$

can be constructed[1,4] in complete analogy with their counterpart in classical statistical mechanics, and are Galilei-covariant and conserved. Moreover, in the case of a configuration space wave functions $\psi(\mathbf{x},t)$ with continuous first partial derivatives in the $\mathbf{x}$-variables, this new quantum probability current converges in the sharp-point limit to the conventional probability current in configuration space,

$$\mathbf{j}_\ell(\mathbf{x},t) \xrightarrow[\ell\to+0]{} (2im)^{-1}\psi^*(\mathbf{x},t)\overleftrightarrow{\nabla}\,\psi(\mathbf{x},t) \ . \tag{2.10}$$

This provides further support to the earlier described physical interpretation of the transition amplitude in Eq. (2.4).

On account of Eq. (2.3), the following family of transition probability amplitudes for free quantum propagation,

$$K_\ell(\mathbf{q}'',\mathbf{p}'',t'';\mathbf{q}',\mathbf{p}',t') \ = \ \left\langle U(t'',\mathbf{q}'',\mathbf{p}''/m,I)\xi^{(\ell)}\middle|U(t',\mathbf{q}',\mathbf{p}'/m,I)\xi^{(\ell)}\right\rangle \ , \tag{2.11}$$

displays the basic properties of a free propagator:

$$K_\ell(\mathbf{q}'',\mathbf{p}'',t'';\mathbf{q}',\mathbf{p}',t') = \int_{\mathbf{R}^6} K_\ell(\mathbf{q}'',\mathbf{p}'',t'';\mathbf{q},\mathbf{p},t)K_\ell(\mathbf{q},\mathbf{p},t;\mathbf{q}',\mathbf{p}',t')\,d\mathbf{q}d\mathbf{p}. \tag{2.12}$$

This phase-space free propagator can be used[4,19] in the formulation of path integrals for quantum propagators in the presence of interactions.

Upon suitable renormalization, these phase space free propagators converge in the sharp-point limit to the well-known free Feynman propagator for nonrelativistic quantum point particles:

$$(\pi/2\ell^2)^{3/2}K_\ell(\mathbf{q}'',\mathbf{p}'',t'';\mathbf{q}',\mathbf{p}',t') \xrightarrow[\ell\to+0]{} K(\mathbf{q}'',t'';\mathbf{q}',t') \ . \tag{2.13}$$

However, the above required renormalization constant obviously diverges in this sharp-point limit. Closer scrutiny reveals[4,19] that the basic formulae,

$$K_\ell(\mathbf{q}'',\mathbf{p}'',t'';\mathbf{q}',\mathbf{p}',t') = \lim_{\varepsilon\to+0}\int K_\ell(\mathbf{q}_N,\mathbf{p}_N,t_N;\mathbf{q}_{N-1},\mathbf{p}_{N-1},t_{N-1})$$
$$\times \prod_{n=N-1}^{1} K_\ell(\mathbf{q}_n,\mathbf{p}_n,t_n;\mathbf{q}_{n-1},\mathbf{p}_{n-1},t_{n-1})\,d\mathbf{q}_n d\mathbf{p}_n \ , \qquad \varepsilon = (t''-t')/N \ , \tag{2.14}$$

on which path integration in general relies, are mathematically well-defined in case of the phase space propagators in Eq. (2.11). On the other hand, in the case of the Feynman propagators which rely on pointlike localization in configuration space, the same type of formulae,

$$K(\mathbf{x}'',t'';\mathbf{x}',t') = \lim_{\varepsilon\to+0}\int K(\mathbf{x}(t_N);\mathbf{x}(t_{N-1})) \prod_{n=N-1}^{1} K(\mathbf{x}(t_n);\mathbf{x}(t_{n-1}))\; d\mathbf{x}(t_n), \quad (2.15)$$

are merely formal, since the above Lebesgue integrals exist if and only if their integrand are integrable in the absolute sense[18] – whereas that is not the case in Eq. (2.15). Hence, the mathematically rigorous treatment of Feynman path integrals require analytic continuations to imaginary time, leading to the well-known Feynman-Kac formula. On the other hand, the phase space path integrals resulting from Eq. (2.14) are mathematically well-defined in *real* time. This constitutes the basis of the usefulness of their special relativistic counterparts, described in the next section. Moreover, it presents decided advantages in the general relativistic regime, which will be our main concern in these lecture notes.

## 3. Special Relativistic Geometro-Stochastic Quantization

In the special relativistic regime the role played in geometro-stochastic quantization by the Galilei group is taken over by the Poincaré group. Similarly, that of nonrelativistic phase space is taken over by a relativistic phase space whose elements are labelled by the 4-vectors $q$ and $p = mv$, with $p$ restricted to the forward and backward mass hyperboloids in the case of particles and antiparticles, respectively. Hence, in the case of particles of rest mass $m$ and zero spin, the Hilbert space $L^2(\Gamma)$ of Sec. 2 is replaced with the Hilbert space $L^2(\Sigma)$ of wave functions $\varphi(q,v)$ which are solutions in $q$ of the free Klein-Gordon equation for rest mass $m$ at each fixed value of $v$ on the forward 4-velocity hyperboloid $\boldsymbol{V}^+$, and whose inner product is provided by the following integral over a hypersurface $\Sigma = \sigma \times \boldsymbol{V}^+$ in relativistic phase space

$$\langle\varphi|\varphi'\rangle = \int_\Sigma \varphi^*(q,v)\varphi'(q,v)\, d\Sigma(q,v), \qquad d\Sigma(q,v) = 2v_\mu \delta(v^2-1) d\sigma^\mu(q) d^4v \;, \quad (3.1)$$

where the $\sigma$-component of $\Sigma$ is a maximal spacelike hypersurface in Minkowski space. The above integration has to be carried out with respect to the unique (modulo a multiplicative constant) Lorentz covariant measure over $\Sigma$, so that the measure element $m^3 d\Sigma$ reduces to $d\mathbf{q}d\mathbf{p}$ along any hypersurface $\Sigma$ for which $\sigma$ corresponds to a constant value of $q^0$ in the chosen global Lorentz frame of reference $\boldsymbol{u}$.

The nonrelativistic representations in (2.1) is now replaced by a representation which assigns the following operator,

$$\big(U(a,\Lambda)\varphi\big)(q,v) = \varphi(\Lambda^{-1}(q-a), \Lambda^{-1}v) \;. \quad (3.2)$$

to the generic element $(a,\Lambda)$ of the Poincaré group, consisting of a spacetime translation $a$ and Lorentz transformation $\Lambda$. As in the nonrelativistic case, the complete harmonic analysis of this representation reveals a host of irreducible subrepresentations that are unitarily equivalent to well-known momentum-space representations.[14] Each such subrepresentation is uniquely characterized by a resolution generator $\eta$ providing the orthogonal projector of $L^2(\Sigma)$ onto the subspace which carries it, since it generates the generalized coherent states

$$\eta_\zeta = U(q,\Lambda_v)\eta \in L^2(\Sigma)\,, \qquad \zeta = (q,v)\;, \quad (3.3)$$

if $\Lambda_v$ denotes the Lorentz boost to the 4-velocity $v$, and, moreover, that projector can be expressed in the form

$$\mathbf{P}_\eta = \int_\Sigma \left|\eta_\zeta\right\rangle d\Sigma(\zeta) \left\langle \eta_\zeta\right| . \tag{3.4}$$

Hence, by an extrapolation of the interpretation of (2.4) in the nonrelativistic case,

$$\varphi(q,v) = \langle U(q,\Lambda_v)\eta|\varphi\rangle , \qquad \varphi \in \mathbf{P}_\eta L^2(\Sigma) , \tag{3.5}$$

can be interpreted as a probability amplitude of detection of a quantum particle in a state described by the normalized vector $\varphi$ in relation to the constituents of a special relativistic quantum frame associated to the classical Lorentz frame of reference $\boldsymbol{u}$.

We note that the above probability amplitude depends on the mean 4-velocity $v$ of each micro-detector in the quantum frame relative to the corresponding classical frame. This is in complete accord with one of Born's key observations on an underlying reciprocity between spacetime and 4-momentum relationships in nature, as clearly indicated by the following quotation: "Ordinary relativity is based on the invariance of the 4-dimensional distance, or its square $R = x_k x^k$. Can one really define the distance of two particles in sub-atomic dimensions independently of their velocity? This seems to me not evident at all." (Ref. 5, p. 208.)

Born's idea of a quantum metric operator $D^2 = Q^2 + P^2$ containing a fundamental length $\ell$ that restricts the sharpness of spatio-temporal localizability leads (cf. Ref. 4, Sec. 4.5) to the *fundamental* quantum frames whose constituents are described by the generalized coherent states in (3.3), represented by the wave functions

$$\Phi^{\boldsymbol{u}}_{\ell,m;\zeta}(q',v') = \tilde{Z}^{-2}_{\ell,m}\int_{k^0>0} \exp\{[i(q-q') - \ell(v'+v)]\cdot k\}\, \delta(k^2-m^2)\, d^4k \ . \tag{3.6}$$

The normalization constant in Eq. (3.6) can be expressed in terms of the modified Bessel function $\mathrm{K}_2$,

$$\tilde{Z}_{\ell,m} = 8\pi^4\, \mathrm{K}_2(2\ell m)/\ell m^2 = (4\pi^4/\ell^3 m^4) + O(\ell^{-1}) \xrightarrow[\ell\to+0]{} +\infty \ . \tag{3.7}$$

This constant diverges in the sharp-point limit $\ell \to 0$, but its value cannot be arbitrarily adjusted, since it is fixed by the fact that the operator in Eq. (3.4) is an orthogonal projection operator, so that

$$\mathbf{P}^{\boldsymbol{u}}_{\ell,m} = \int \left|\Phi^{\boldsymbol{u}}_{\ell,m;\zeta}\right\rangle d\Sigma(\zeta) \left\langle \Phi^{\boldsymbol{u}}_{\ell,m;\zeta}\right| = \left(\mathbf{P}^{\boldsymbol{u}}_{\ell,m}\right)^* = \left(\mathbf{P}^{\boldsymbol{u}}_{\ell,m}\right)^2 \ . \tag{3.8}$$

As a consequence, no counterparts of Eqs. (2.7) and (2.10) exist. This is in agreement with the fact that, as pointed out by Wigner in a lecture series on conventional nonrelativistic and special relativistic quantum theory, "every attempt to provide a precise definition of a position coordinate [for a quantum point particle] stands in direct contradiction with special relativity" (Ref. 6, p. 313).

Despite this fact, a consistent special relativistic quantum theory of spacetime localization is feasible if a *fixed* value $\ell > 0$ is adopted as a fundamental length, since then the following current, constructed in analogy with the one in Eq. (2.9),

$$j^\mu_\ell(q) = 2\int_{v^0>0} v^\mu \left|\varphi_\ell(q,v)\right|^2 \delta(v^2-1)\, d^4v , \qquad \varphi_\ell(q,v) = \left\langle \Phi^{\boldsymbol{u}}_{\ell,m;q,v}\middle|\varphi\right\rangle \ , \tag{3.9}$$

is Poincaré-covariant and conserved. Most importantly, it has a positive-definite timelike component, so that it provides a *bona fide* relativistic *probability* current.

This does not mean that the usual treatment of the Klein-Gordon equation has to be abandoned in the present framework, since the following counterpart,

$$J^\mu_\ell(q) = i\hat{Z}_{\ell,m}\int_{v^0>0} \varphi^*_\ell(q,v)\overleftrightarrow{\partial}^\mu \varphi_\ell(q,v)\, \delta(v^2-1)\, d^4v , \qquad \partial_\mu = \partial/\partial q^\mu \ , \tag{3.10}$$

of the familiar Klein-Gordon current also exists, and it is Poincaré-covariant as well as conserved. However, as it is the case in the conventional treatment of the Klein-Gordon equation, this latter current is *not* a probability current, since its timelike component is *not* positive definite even for wave functions corresponding to positive energies. Hence, as in the conventional approach, its use is restricted to the formulation of quantum field theoretical couplings of Klein-Gordon fields to other fields in the quantum-geometric field theory discussed in subsequent sections.

The normalization constant in Eq. (3.10) is uniquely determined by the following alternative form,[2,4]

$$\left\langle \varphi_\ell \middle| \varphi'_\ell \right\rangle = i\hat{Z}_{\ell,m} \int_{v^0>0} \varphi^*_\ell(q,v)\overleftrightarrow{\partial}_\mu \varphi'_\ell(q,v)\, d\sigma^\mu(q)\, \delta(v^2-1) d^4v \quad , \qquad (3.11)$$

of the restriction of the inner product in (3.1) to subspace of $L^2(\Sigma)$ onto which the projector in Eq. (3.8) projects. Its value is finite for all $\ell > 0$, but it diverges in the sharp-point limit:

$$\hat{Z}_{\ell,m} = \mathrm{K}_2(2\ell m)/m\mathrm{K}_1(2\ell m) = (1/\ell m^2) + O(1) \xrightarrow[\ell\to+0]{} +\infty \quad . \qquad (3.12)$$

As shown in Refs. 9 and 10, the divergence in the sharp-point limit $\ell \to 0$ of the two normalization constants in Eqs. (3.7) and (3.12) provides an explanation of the root causes of the divergences that plague the conventional formulation of QFT. In essence, this explanation is that, as maintained already in the 1930s by Bohr, Born, Dirac, Heisenberg, and other founders of quantum mechanics, there *does* exist a fundamental length in nature. Hence, the appearance of all the divergences that plague conventional QFT is simply a manifestation of this underlying fact.

In the present context, as a consequence of (3.8), we have

$$K_\ell(q'',v'';q',v') = \left\langle \boldsymbol{\Phi}^{\boldsymbol{u}}_{q'',v''} \middle| \boldsymbol{\Phi}^{\boldsymbol{u}}_{q',v'} \right\rangle = \int K_\ell(q'',v'';q,v) K_\ell(q,v;q',v')\, d\Sigma(q,v) . \quad (3.13)$$

Hence, as was the case in Eqs. (2.11) and (2.12), the above transition probability amplitudes display all the basic properties of a propagator, and can be used[4,19] in the formulation of path integrals. On the other hand, in the sharp-point limit $\ell \to +0$ we can carry out their renormalization, so as to obtain,

$$2(2\pi)^{-3} Z^2_\ell K_\ell(q'',p'';q',p') \xrightarrow[\ell\to+0]{} K_F(q''-q') \quad , \qquad q''^0 > q'^0 \, , \qquad (3.14)$$

i.e., so that the forward time-ordered phase space propagator determined by (3.13) converges in that sharp-point limit to the corresponding forward part of the time-ordered Feynman propagator for Klein-Gordon particles, which is given by

$$K_F(x''-x') = 2(2\pi)^{-3} \int_{k^0>0} \exp[ik\cdot(x'-x'')]\, \delta(k^2-m^2) d^4k \, , \qquad x''^0 > x'^0 \, . \quad (3.15)$$

This basic feature contributes to the *formal* term-by-term convergence of formal perturbative expansions of transition amplitudes for collision processes in quantum-geometric field theory to their conventional counterparts in models involving Klein-Gordon quantum fields. Such term-by-term convergence can be achieved after other factors that diverge in the sharp-point limit have also been discarded – namely those divergent factors that appear in the formal sharp-point limit due to the independence of free Feynman propagators on the stochastic 4-velocity variable $v$ .[9,10,17]

Totally analogous comparisons between the quantum-geometric and the conventional approach to QFT can be drawn[9,10] in the case of the Dirac equation, quantum electromagnetic fields, etc. However, in the presence of gravity, the retention of a fixed value $\ell > 0$ for the fundamental length becomes mandatory, since very basic measurement-theoretical considerations[20,21] show that Planck's length imposes an absolute lower bound on $\ell$, so that no spacetime localization can exceed the precision stipulated by Planck's length.

## 4. Quantum Geometry and General Relativity

The basic physical idea in extending the geometro-stochastic method of quantization to the semiclassical general relativistic regime in which a Lorentzian manifold $(\mathbf{M},\boldsymbol{g}^{\mathrm{L}})$ is assumed to be *a priori* given, is based on the observation that, due to the strong equivalence principle, in classical general relativity (CGR) the role played by the Minkowski space $M^4 \cong (\mathbf{R}^4,\eta)$ is taken over by the fibres $T_x\mathbf{M}$ of the tangent bundle $T\mathbf{M}$, whose typical fibre is $\mathbf{R}^4$; moreover, the role of the global Lorentz frames $\mathcal{L} = \{\boldsymbol{e}_i(x_0) \mid i = 0,1,2,3\}$, with their origins at points $x_0 \in M^4$, is taken over by the local Lorentz frames within the Lorentz frame bundle $L\mathbf{M}(\boldsymbol{g}^{\mathrm{L}})$, which has the Lorentz group SO(3,1) as its structure group. Hence, all the tensor fields that appear in CGR represent sections of various tensor bundles associated with $L\mathbf{M}(\boldsymbol{g}^{\mathrm{L}})$. The resulting Lorentz gauge invariance is underlined by the fact that the Levi-Civita connection employed in CGR is compatible with the metric $\boldsymbol{g}^{\mathrm{L}}$.

Mathematically, this Lorentz gauge invariance can be always extended into Poincaré gauge invariance based on the Poincaré frame bundle $P\mathbf{M}(\boldsymbol{g}^{\mathrm{L}})$, that has the Poincaré group ISO(3,1) as its structure group. The Levi-Civita connection on $L\mathbf{M}(\boldsymbol{g}^{\mathrm{L}})$ is then extended into a connection on $P\mathbf{M}(\boldsymbol{g}^{\mathrm{L}})$, so that, as shown by Drechsler[22], the operator forms $\nabla$ for covariant differentiation acquire four additional terms contained in the connection 1-forms

$$\tilde{\boldsymbol{\theta}}^i = \boldsymbol{\theta}^i + (\nabla \boldsymbol{a})^i \ , \qquad \boldsymbol{s} : x \mapsto (\boldsymbol{a},\boldsymbol{e}_i) \ , \quad \boldsymbol{a} = a^i \boldsymbol{e}_i \ , \tag{4.1}$$

for any section $\boldsymbol{s}$ of $P\mathbf{M}(\boldsymbol{g}^{\mathrm{L}})$ – where the canonical forms $\theta^i$ are provided by the coframes $\{\theta^i \mid i = 0,1,2,3\}$ dual to the Lorentz frames $\{\boldsymbol{e}_i \mid i = 0,1,2,3\}$ in $\boldsymbol{s}$; and where each Poincaré frame $\{(\boldsymbol{a},\boldsymbol{e}_i) \mid i = 0,1,2,3\}$ is obtained by translating the corresponding local Lorentz frame by the 4-vector $\boldsymbol{a}$ within the tangent space within which that frame lies.

Such an extension is not esential to CGR itself, but it is essential from the quantum point of view. This is due to the fact that the kinematics and dynamics incorporated into such equations of motion as those of Klein-Gordon, Dirac, etc., which govern quantum propagation, gives rise to generators of infinitesimal spacetime translations for unitary representations of the Poincaré group.

Hence, in order to describe quantum-geometrically the propagation of massive spin-0 quantum fields, we shall construct the Fock-Klein-Gordon bundle

$$\mathcal{E}(\mathbf{M},\boldsymbol{g}^{\mathrm{L}}) = P\mathbf{M}(\boldsymbol{g}^{\mathrm{L}}) \times_{\mathbf{G}} \mathcal{F} \ , \qquad \mathbf{G} = \mathrm{ISO}(3,1) \ , \tag{4.2}$$

associated with the principal bundle $P\mathbf{M}(\boldsymbol{g}^{\mathrm{L}})$. The typical fibre of this quantum bundle is the Fock space

$$\mathcal{F} = \bigoplus_{n=0}^{\infty} \mathcal{F}_n \ , \qquad \mathcal{F}_n = \mathbf{F} \underset{S}{\otimes} \cdots \underset{S}{\otimes} \mathbf{F} \ , \qquad \mathbf{F} = \mathbf{P}^{\boldsymbol{u}_0}_{\ell,m} L^2(\Sigma) \ , \tag{4.3}$$

obtained from symmetrized tensor products of the Hilbert space determined for the standard frame $\boldsymbol{u}_0$ in $(\mathbf{R}^4,\eta)$ by Eqs. (3.5), (3.6) and (3.8). Hence, the representation provided by Eq (3.2) gives rise within this Fock space to the representation

$$\boldsymbol{U}(a,\Lambda) = \bigoplus_{n=0}^{\infty} U(a,\Lambda)^{\otimes n} \ , \qquad (a,\Lambda) \in \mathrm{ISO}(3,1) \ , \tag{4.4}$$

which can be used in the standard manner in the construction of the bundle $\mathbf{G}$-product[9,10] that appears in Eq. (4.2), so that the Poincaré group becomes the structure group of $\mathcal{E}(\mathbf{M},\boldsymbol{g}^{\mathrm{L}})$. We observe that the fibre $\mathcal{F}_x$ above $x \in \mathbf{M}$ contains the vacuum subfibre $\mathcal{F}_{0;x}$ spanned by a *local* Fock vacuum state vector $\Psi_{0;x}$, which is left invariant by the counterpart of Eq. (4.4) in $\mathcal{F}_x$.

The operational description of measurements of spatio-temporal separations of events in a neighborhood of some base point $x \in \mathbf{M}$ where the relative curvature effects are small, so that classically they can be described exclusively in terms of Riemann normal coordinates for which $dx_k dx^k = \eta_{ik} dx^i dx^k$ at $x \in \mathbf{M}$, can be now supplemented,[10] in accordance with Born's[5] orginal physical and epistemological ideas, by fundamental uncertainties described by the local quantum fluctuation amplitudes

$$\Delta_x^{(+)}(\zeta';\zeta) = -i\ \boldsymbol{\Phi}^{\boldsymbol{u}(x)}_{\ell,m;\zeta}(\zeta'), \quad \boldsymbol{\zeta} = (\boldsymbol{a} + q^i \boldsymbol{e}_i, v^i \boldsymbol{e}_i) \in T_x\mathbf{M} \times \boldsymbol{V}_x^+ , \quad \zeta = q - i\ell v \in \mathbf{C}^4 . \tag{4.5}$$

According to the interpretation of Eq. (3.5), these fluctuations result from the transition probability amplitudes of the geometrically[10] *local quantum frames* associated by the construction in Eq. (4.2) to each Poincaré frame $\boldsymbol{u}(x) = (\boldsymbol{a}(x), \boldsymbol{e}_i(x))$, that can be used to define those Riemann normal coordinates by means of the exponential map at $x$.

On account of Eqs. (3.8) and (3.11), we have

$$2\int \Delta_x^{(+)}(\zeta';\zeta)\, \Delta_x^{(+)}(\zeta;\zeta'')\, \delta(v^2-1)\, v_k\, d\sigma^k(q) d^4 v = -i \Delta_x^{(+)}(\zeta';\zeta'') , \tag{4.6}$$

so that the 2-point functions in Eq. (4.5) can also play the role of geometrically local propagators. Hence, with their help we can define annihilation operators by their action upon $n$-exciton state vectors $\Psi_{n;x} \in \mathcal{F}_{n;x}$ as follows:

$$\left(\varphi^{(-)}(x;\zeta)\boldsymbol{\Psi}_{n;x}\right)_{n-1}(\zeta_1,\ldots,\zeta_{n-1}) = i n^{1/2} \int \Delta_x^{(+)}(\zeta;\zeta_n)\, \boldsymbol{\Psi}_{n;x}(\zeta_1,\ldots,\zeta_n)\, d\Sigma(\zeta_n) . \tag{4.7}$$

In conventional QFT the counterparts of these operators, and of their adjoints representing creation operators, have to be smeared with test functions. Indeed, in that context such operators are *proven* to exist only in the sense of operator-valued distributions (cf. Ref. 23, Sec. 10.4). On the other hand, the presence of the fundamental length $\ell$ in the 2-point functions in Eq. (4.5) renders them non-singular. This makes the annihilation operators in Eq. (4.7), as well as of their adjoints $\varphi^{(+)}(x;\zeta)$ representing creation operators, mathematically well-defined without any need for “smearing” with test functions. Hence, the Klein-Gordon *quantum frame field*

$$\varphi(x;\zeta) = \varphi^{(+)}(x;\zeta) + \varphi^{(-)}(x;\zeta) , \quad \zeta \in T_x\mathbf{M} \times \boldsymbol{V}_x^+ , \quad x \in \mathbf{M} , \tag{4.8}$$

is well-defined as it stands. As a consequence, it can be proved[8-10] that when the Poincaré group acts upon the elements $\boldsymbol{s}(x)$ of a section $\boldsymbol{s}$ of $P\mathbf{M}(\boldsymbol{g}^{\mathrm{L}})$, the infinitesimal generators of the counterparts

$$\boldsymbol{U}_{\boldsymbol{s}(x)}(a,\Lambda) = \bigoplus_{n=0}^{\infty} U_{\boldsymbol{s}(x)}(a,\Lambda)^{\otimes n} , \qquad \boldsymbol{s}(x) \in P\mathbf{M}(\boldsymbol{g}^{\mathrm{L}}) , \tag{4.9}$$

of the representation in Eq. (4.4) are given by the well-defined Bochner integrals (cf. Ref. 18, p. 480)

$$\boldsymbol{P}_{j;\boldsymbol{s}(x)} = \int : T_{jk}[\varphi(x;\zeta)] : d\sigma^k(q)\, \delta(v^2-1)\, d^4 v , \quad Q^j_{\boldsymbol{s}(x)} = q^j - (i/m)\partial/\partial v_j , \tag{4.10}$$

$$\boldsymbol{M}^{ij}_{\boldsymbol{s}(x)} = \int : Q^i_{\boldsymbol{s}(x)} T^{jk}[\varphi(x;\zeta)] - Q^j_{\boldsymbol{s}(x)} T^{ik}[\varphi(x;\zeta)] : d\sigma_k(q)\, \delta(v^2-1)\, d^4 v , \tag{4.11}$$

whose integrands are *bona fide* operator-valued functions, given by the normally-ordered values of the following *bona fide* operator-valued stress-energy tensor:

$$T_{jk}[\varphi] = \hat{Z}_{\ell,m}\Big(\varphi_{,j}\varphi_{,k} + \tfrac{1}{2}\eta_{jk}\big(m^2\varphi^2 - \eta^{il}\varphi_{,i}\varphi_{,l}\big)\Big), \quad \varphi_{,j} = \partial\varphi/\partial q^j \ . \tag{4.12}$$

However, as can be seen from Eq. (3.12), divergencies manifest themselves in the sharp-point limit $\ell \to +0$.

In each Fock fibre $\mathcal{F}_x$ we can introduce the Glauber-type coherent states

$$\boldsymbol{\Phi}_{\mathbf{f}} = \exp\Big[-\tfrac{1}{2}\langle\mathbf{f}|\mathbf{f}\rangle + \varphi^{(+)}(\mathbf{f})\Big]\,\boldsymbol{\Psi}_{0;x}, \quad \varphi^{(+)}(\mathbf{f}) = \int \varphi^{(+)}(x;\boldsymbol{\zeta})\mathbf{f}(\boldsymbol{\zeta})\,d\Sigma(\boldsymbol{\zeta}), \quad \mathbf{f} \in \mathcal{F}_{1;x}, \tag{4.13}$$

defined by the strongly convergent power series for the above exponential. These coherent states constitute a family of second-quantized frames, in the sense that they give rise to the following continuous resolution of the identity operator $\mathbf{1}_x$ in each fibre $\mathcal{F}_x$ ,

$$\int_{\mathbf{F}_x} \big|\boldsymbol{\Phi}_{\mathbf{f}}\big\rangle\, \mathbf{df}\,\mathbf{df}^*\big\langle\boldsymbol{\Phi}_{\mathbf{f}}\big| \;=\; \mathbf{1}_x \ , \qquad \mathbf{F}_x = \mathcal{F}_{1;x} \ , \tag{4.14}$$

where the above functional integral is mathematically rigorously defined by the method of Berezin.[25] These continuous resolutions of the identity will prove essential in Sec. 5) to the derivation of action integrals for quantum-geometric propagation, when taken in conjunction with the fact that the elements of these second-quantized frames are *bona fide* eigenvectors of the annihilation operators in Eq. (4.7):

$$\varphi^{(-)}(x;\boldsymbol{\zeta})\boldsymbol{\Phi}_{\mathbf{f}} = \mathbf{f}(\boldsymbol{\zeta})\,\boldsymbol{\Phi}_{\mathbf{f}} \ , \qquad \mathbf{f} \in \mathbf{F}_x \ . \tag{4.15}$$

This quantum-geometric propagation emerges from the fact that the quantum bundle $\mathcal{E}(\mathbf{M},\boldsymbol{g}^{\mathrm{L}})$ is associated with $P\mathbf{M}(\boldsymbol{g}^{\mathrm{L}})$, and consequently each connection on $P\mathbf{M}(\boldsymbol{g}^{\mathrm{L}})$ gives rise to parallel transport within $\mathcal{E}(\mathbf{M},\boldsymbol{g}^{\mathrm{L}})$. In particular, this is the case with the earlier discussed extension to $P\mathbf{M}(\boldsymbol{g}^{\mathrm{L}})$ of the Levi-Civita connection on $L\mathbf{M}(\boldsymbol{g}^{\mathrm{L}})$. Therefore, for any choice of Poincaré gauge given by a section $\boldsymbol{s}$ of $P\mathbf{M}(\boldsymbol{g}^{\mathrm{L}})$, the corresponding parallel transport determines the quantum connection[7-10] (cf. Ref. 38 about the sign conventions)

$$\boldsymbol{\nabla} = \boldsymbol{d} + i\tilde{\boldsymbol{\theta}}^i \boldsymbol{P}_{i;\boldsymbol{s}} + \tfrac{i}{2}\tilde{\boldsymbol{\omega}}_{jk}\boldsymbol{M}_{\boldsymbol{s}}^{jk} \ , \qquad \boldsymbol{d} = \boldsymbol{\theta}^i \boldsymbol{\partial}_{\boldsymbol{e}_i} \ , \tag{4.16}$$

whose connection 1-forms are those derived from that extension, and whose infinitesimal generators for spacetime translations and Lorentz transformations are the ones in Eqs. (4.10) and (4.11). Thus, Poincaré gauge covariance is embedded into the parallel transport resulting from this above quantum connection.

## 5. Quantum-Geometric Propagation in Quantum Bundles

In a stationary but nonstatic classical spacetime represented by a Lorentzian manifold $(\mathbf{M},\boldsymbol{g}^{\mathrm{L}})$, the conventional QFT formulation of free-fall field propagation gives rise to *ex nihilo* particle production.[26] This violates local energy-momentum conservation as well as the strong equivalence principle of general relativity since, according to that fundamental principle, all observers in free fall are *inertial*; consequently, the situation should appear to them *locally* the same as in Minkowski space, where such *ex nihilo* particle production is not in evidence.

In the quantum-geometric formulation of QFT, this difficulty can be overcome by applying to all the fields defined on quantum bundles over curved spacetime the *original* form of the path-integration method advanced by Feynman. In that original form[27,28] the path-integral depiction of quantum propagation was formulated in spacetime, rather than in the nowadays more popular momentum representation. Its two central ideas were that such propagation takes place over broken polygonal paths in nonrelativistic or relativistic flat spacetimes, and that all the observed probability transition amplitudes result from the superposition of *spacetime* probability amplitudes over such paths.

The quantum-geometric adaptation[7-10] of these ideas to a curved spacetime represented by a Lorentzian manifold $(\mathbf{M},\boldsymbol{g}^{\mathrm{L}})$ is based on replacing the straight lines of such broken polygonal paths with the arcs of geodesics of the Lorentzian metric $\boldsymbol{g}^{\mathrm{L}}$, and on deriving spacetime propagators by extending the strong equivalence principle from the classical to the quantum regime. The latter feature entails that free-fall quantum-geometric propagation is governed by the Levi-Civita connection determined by $\boldsymbol{g}^{\mathrm{L}}$, so that *ex nihilo* particle production is thereby avoided. Moreover, a central role in such quantum-geometric evolution is played by the *local* proper energy density and 3-momentum of the nongravitational sources associated, as is the case in classical canonical gravity, with a geometrodynamic evolution of spacetime that is mathematically equivalent to a foliation of Lorentzian manifold $(\mathbf{M},\boldsymbol{g}^{\mathrm{L}})$ into a family of reference hypersurfaces $\Sigma_t$ with unit future-pointing normals $\boldsymbol{n}(x) \in T_x\mathbf{M}$.

The presence of the fundamental length $\ell$ makes the definition of such geometrically *local* quantities not only mathematically consistent, but also in keeping with the uncertainty principle, since the envisaged spacetime localization is not infinitely precise, but exhibits stochastic fluctuations described by the quantum fluctuations amplitudes in Eq. (4.5). Hence, let us introduce the quantum-geometric field

$$\boldsymbol{\varphi}(x,\boldsymbol{v}) := \varphi(x;(\mathbf{0},\boldsymbol{v})) \quad , \qquad \boldsymbol{v} = v^i \boldsymbol{e}_i(x) \in \boldsymbol{V}_x^+ \ , \tag{5.1}$$

whose values are actually independent of the chosen local Poincaré frame $(\boldsymbol{a}(x),\boldsymbol{e}_i(x)) \in P\mathbf{M}(\boldsymbol{g}^{\mathrm{L}})$; furthermore, let us consider for any section $\boldsymbol{s} = \{\boldsymbol{a}(x),\boldsymbol{e}_i(x) | \ x \in \mathbf{M}\}$ the local energy-momentum density operators

$$\hat{\boldsymbol{P}}_{j;\boldsymbol{s}}(x) = n^k(x) \int_{v^0>0} :T_{jk}[\boldsymbol{\varphi}(x;\boldsymbol{v})]: \ \delta(v^2-1)\, d^4v \ , \tag{5.2}$$

where $n^k(x)$, $k = 0,1,2,3$, are the components of the normal $\boldsymbol{n}(x)$ with respect to the local Lorentz frame $\{\boldsymbol{e}_i(x)\}$, $x \in \Sigma_t$, and the stress-energy operator is the one that appears in Eq. (4.12). If the Poincaré gauge represented by $\boldsymbol{s}$ is adapted to the family of reference hypersurface $\Sigma_t$, in the sense that the timelike components of its vierbeins are orthogonal to those hypersurfaces, then the proper energy operator $\rho$ and the 3-momentum operators $\boldsymbol{j}_a$ of the quantum-geometric field $\varphi$ can be defined as in classical canonical gravity:

$$\boldsymbol{\rho}(x) = \hat{\boldsymbol{P}}_{0;\boldsymbol{s}}(x) \ , \qquad \boldsymbol{j}_a(x) = \hat{\boldsymbol{P}}_{a;\boldsymbol{s}}(x) \ , \qquad a = 1,2,3 \ . \tag{5.3}$$

We note that, as opposed to the quantum frame field in Eq. (4.8), the quantum-geometric field in Eq. (5.2) does not depend on the Poincaré gauge variables $q^j$, $j = 0,1,2,3$. The reason is that, although these variables are required for the definition of infinitesimal spacetime translations at each base location $x \in \mathbf{M}$, they cannot play a direct physical role, as it becomes evident as soon as the transition to the special relativistic regime is performed. Indeed, in case that $(\mathbf{M},\boldsymbol{g}^{\mathrm{L}})$ is the Minkowski space $M^4 \cong (\mathbf{R}^4,\eta)$, a global Lorentz frame $\mathcal{L} = \{\boldsymbol{e}_i(x_0) | \ i = 0,1,2,3\}$ with its origin at $x_0 \in M^4$ corresponds to the following cross-section of the Poincaré frame bundle $PM^4$,

$$\boldsymbol{s}_0(\mathcal{L}) = \left\{ (\boldsymbol{a}(x),\boldsymbol{e}_i(x)) \middle| \ \boldsymbol{a}(x) = -x^i \boldsymbol{e}_i(x) \in T_x M^4 \, , \ x = x^i \boldsymbol{e}_i(x_0) \in M^4 \right\} \ , \tag{5.4}$$

due to the fact that each tangent space $T_x M^4$ can be identified with $M^4$ itself. Hence, each special-relativistic state vector $\varphi \in \mathbf{F}$ can be then identified with a cross-section $\{ \Psi_{1;x} \in \mathcal{F}_x | \ x$ $\in \mathbf{M}\}$ of the quantum bundle $\mathcal{E}(M^4)$ whose local state vectors $\Psi_{1;x}$ have coordinate wave functions that satisfy

$$\Psi_{1;x}(-i\ell v) = \varphi(-a(x) - i\ell v) \ , \qquad -a(x) = (x^0,\ldots,x^3) \in \mathbf{R}^4, \tag{5.5}$$

in the Poincaré frames of $\boldsymbol{s}_0(\mathcal{L})$; by analytic continuation the values of $\Psi_{1;x}(q - i\ell v)$ are then uniquely determined at all $q \in \mathbf{R}^4$. In view of the Poincaré gauge invariance of the quantum geometry framework, this identification is invariant under changes of global Lorentz frames $\mathcal{L}$, since all such changes induce corresponding changes in the global Poincaré gauges of the type in Eq. (5.4). Hence, the special relativistic framework of Sec. 3 is recovered as a special case of the present Poincaré gauge invariant framework, and the gauge variables $q^j$ are seen to be physically redundant.

This suggests the construction of the Fock space

$$\hat{\mathcal{F}} = \bigoplus_{n=0}^{\infty} \hat{\mathcal{F}}_n \ , \qquad \hat{\mathcal{F}}_n = \hat{\mathbf{F}} \underset{S}{\otimes} \cdots \underset{S}{\otimes} \hat{\mathbf{F}} \ , \qquad \hat{\mathbf{F}} = \left\{ \hat{f} \,\middle|\, f \in \mathbf{F} \right\} \ , \tag{5.6}$$

whose state vectors consist of $n$-exciton field modes $\hat{\Psi}_n \in \hat{\mathcal{F}}_n$ that are related to the corresponding $\Psi_n \in \mathcal{F}_n$ by the equalities

$$\hat{\Psi}_n(v_1,\ldots,v_n) = \Psi_n(-a(x) - i\ell v_1,\ldots,-a(x) - i\ell v_n) \ , \qquad \Psi_n \in \mathcal{F}_n \ . \tag{5.7}$$

By analytic continuation, these equalities establish isomorphisms between $\hat{\mathcal{F}}_n$ and $\mathcal{F}_n$ for $n = 1,2,\ldots$. In turn, these isomorphisms become isometries if the restrictions of Eq. (3.2) to Lorentz transformations are employed to introduce in $\hat{\mathbf{F}}$ Perelomov-type generalized coherent states as in Eq. (3.3). We then have that

$$\langle \hat{\varphi}_1 | \hat{\varphi}_2 \rangle = Z_{\ell,m} \int_{v^0 > 0} \hat{\varphi}_1^*(v) \hat{\varphi}_2(v)\, \delta(v^2 - 1)\, d^4 v = \langle \varphi_1 | \varphi_2 \rangle, \quad \hat{\varphi}_1, \hat{\varphi}_2 \in \hat{\mathbf{F}} \ , \quad \varphi_1, \varphi_2 \in \mathbf{F}, \tag{5.8}$$

where the value of the renormalization constant $Z_{\ell,m}$ can be determined[10,17] by the method presented in Ref. 15.

We can now define a bundle of localized quantum field excitation modes of $\varphi$,

$$\hat{\mathcal{E}}(\mathbf{M}, \boldsymbol{g}^{\mathrm{L}}) = L\mathbf{M}(\boldsymbol{g}^{\mathrm{L}}) \times_{\mathbf{G}} \hat{\mathcal{F}} \ , \qquad \mathbf{G} = \mathrm{SO}(3,1) \ , \tag{5.9}$$

which has the Lorentz group as its structure group. On account of (5.7) its Lorentz gauge covariance can be extended into the Poincaré gauge covariance of the Fock-Klein-Gordon bundle in Eq. (4.2), so that the parallel transport in the latter can be transferred to the former. We can use this fact to define the quantum-geometric evolution between two such hypersurfaces $\Sigma_{t'}$ and $\Sigma_{t''}$ of a geometrodynamic spacetime evolution given by a foliation of the Lorentzian manifold $(\mathbf{M}, \boldsymbol{g}^{\mathrm{L}})$ into synchronous reference hypersurfaces $\Sigma_t$ labeled by a globally defined parameter $t$ .

Let us therefore consider for some integer $N$ the family of reference hypersurfaces $\Sigma_{t_n}$, $n = 1,\ldots,N$, which are such that $\Sigma_{t'} = \Sigma_{t_0}$ and $\Sigma_{t''} = \Sigma_{t_N}$ , and that $t_n - t_{n-1} = (t'' - t')/N = \varepsilon$ . Let us denote by $\mathbf{S}_n$ the segment in the base manifold $(\mathbf{M}, \boldsymbol{g}^{\mathrm{L}})$ between the hypersurfaces $\Sigma_{t_{n-1}}$ and $\Sigma_{t_n}$, to which we shall refer as its inflow and outflow hypersurface, respectively. For a choice of cross-section $\boldsymbol{s} = \{\boldsymbol{a}(x), \boldsymbol{e}_i(x) | \, x \in \mathbf{M}\}$ of the Poincaré frame bundle $P\mathbf{M}$ we then introduce above each one of the inflow-outflow hypersurfaces corresponding to $n = 1,\ldots,N-1$ the coherent field-mode sections

$$\hat{\Phi}^{\boldsymbol{s}}_{\varphi_n(x_n)} \in \hat{\mathcal{F}}_{x_n} \ , \qquad \varphi_n(x_n) \in \mathbf{F}_{x_n} \ , \qquad \varphi_n \in \mathbf{F} \ , \qquad x_n \in \Sigma_{t_n} \ , \tag{5.10}$$

determined[10] from single-boson states $\varphi_n$ within the typical fibre $\mathbf{F}$ in a manner which naturally extrapolates the construction in Eq. (5.5) from flat to curved spacetimes. Let us then consider all the broken paths between $\Sigma_{t'}$ and $\Sigma_{t''}$ which consist of geodesic arcs $\gamma(x_{n-1},x_n)$ connecting points $x_{n-1}$ on inflow hypersurfaces $\Sigma_{t_{n-1}}$ with points $x_n$ on the respective outflow hypersurfaces $\Sigma_{t_n}$ of the base-segments $\mathbf{S}_n$, $n = 1,\ldots,N$.

As mentioned earlier, quantum-geometric propagation proceeds by parallel transport that abides by the strong equivalence principle and by the geodesic postulate. Consequently, if $\tau_\gamma(x_n,x_{n-1})$ denotes the operator for parallel transport along $\gamma(x_{n-1},x_n)$ determined by the quantum connection in Eq. (4.16), the quantum-geometric free-fall propagator is given by

$$K(\varphi(x'');\varphi(x')) = \lim_{\varepsilon\to+0} \prod_{n=N}^{1}{}' \int \mathcal{D}\varphi_n \left\langle e^{i\varepsilon\int d\sigma(x_n)\,\boldsymbol{\rho}(x_n)}\ \hat{\boldsymbol{\Phi}}^{\boldsymbol{s}}_{\varphi_n(x_n)} \middle|\ \tau_\gamma(x_n,x_{n-1})\, \hat{\boldsymbol{\Phi}}^{\boldsymbol{s}}_{\varphi_{n-1}(x_{n-1})} \right\rangle, \quad (5.11)$$

where $\rho(x_n)$ is given by Eqs. (5.1)-(5.3), and $d\sigma(x_n)$ is the Riemannian measure element determined by the 3-metric induced by the Lorentzian metric $\boldsymbol{g}^{\mathrm{L}}$ on the spacelike reference hypersurface $\Sigma_{t_n}$. The above functional integration is to be carried out over all the coherent field modes above $\Sigma_{t_n}$ with respect to the functional "measure"

$$\mathcal{D}\varphi_n = \prod\nolimits_{x_n\in\Sigma_{t_n}} \mathcal{D}[\varphi_n(x_n)] \ , \quad (5.12)$$

in the sense of Riemannian[10] rather than of Lebesgue integration,[18] so that no transition to imaginary time and the "Euclidean regime" is required – as in the case with the Feynman-Kac formula. Rather, the smoothness of coherent sections secures the existence of such integrals over physical families of stochastic paths in *real* spacetime.

The key formula (4.15), together with other properties of coherent states, enables the recasting of Eq. (5.11) in the form of the path integral

$$K(\varphi(x'');\varphi(x')) = \int \mathrm{D}\varphi \exp[iS(\varphi)] \ , \qquad \mathrm{D}\varphi = \prod\nolimits_{t''>t>t'} \prod\nolimits_{x\in\Sigma_t} \mathcal{D}\,\varphi(x) \ , \quad (5.13)$$

based on an action that can be expressed terms of a Lagrangian density:

$$S(\varphi) = \int_{t'}^{t''} dt \int_{\Sigma_\tau} d\sigma(x) \int \delta(v^2-1)\, d^4v\, \mathcal{L}_0(\varphi_t(x,v), \varphi_{t-0}(x,v)) \ , \quad (5.14)$$

$$\mathcal{L}_0 = \tfrac{i}{2} Z_{\ell,m}\Big[\overline{\varphi}_t(x,v)\,\dot{\varphi}_t(x,v) - \dot{\overline{\varphi}}_t(x,v)\,\varphi_{t-0}(x,v)\Big] - T_{00}[\overline{\varphi}_t(x,v)+\varphi_{t-0}(x,v)] \ . \quad (5.15)$$

Thus, in quantum-geometric field theory Lagrangians and action integrals are *derived* from geometric physical principles, rather than being postulated. Nevertheless, it has been shown[10,17] that field interactions can be incorporated without difficulty in the quantum-geometric framework, and that in the special relativistic regime agreement with conventional QFT can be achieved by taking the sharp-point limit $\ell \to 0$ in a *formal* perturbation expansion.

## 6. The Physical Foundations of Geometric Quantum Gravity

The extrapolation of quantum-geometric propagation to quantum gravity, where matter fields are in *mutual* interaction with a quantum gravitational field $\mathbf{g}^{\mathrm{Q}}$, has to take into account the fact that already in classical gravity such mutual interactions between "matter" fields (including those describing nongravitational radiation) and gravitational fields display diffeomorphism invariance.[29] Thus, if the Lorentzian manifold $(\mathbf{M},\boldsymbol{g}^{\mathrm{L}})$ provides a CGR model for spacetime, and if $\psi : \mathbf{M} \to \mathbf{M}$ is any diffeomorphism of $\mathbf{M}$ onto itself, then the Lorentzian manifold $(\mathbf{M},\boldsymbol{g}'^{\mathrm{L}})$ provides a *physically* equivalent model if we set,

$$\boldsymbol{g}'^{\mathrm{L}} = \psi_* \boldsymbol{g}^{\mathrm{L}} \ , \quad \boldsymbol{T}'_\alpha = \psi_* \boldsymbol{T}_\alpha \ , \quad x'_\beta(t) = \psi(x_\beta(t)) \ , \tag{6.1}$$

for all the matter and radiation tensor fields $\boldsymbol{T}_\alpha$, as well as for all causal test-particle world-lines $\{x_\beta(t) \in \mathbf{M}\}$ of that model. Hence, if $Riem^{\mathrm{L}}\mathbf{M}$ is the family of all Lorentzian metrics in $\mathbf{M}$, and if $Diff\mathbf{M}$ is the diffeomorphism group on $\mathbf{M}$, then a CGR model in $\mathbf{M}$ is actually represented by an equivalence class $\mathcal{g}^{\mathrm{L}}$,

$$\mathcal{g}^{\mathrm{L}} = \{ \psi_* \boldsymbol{g}^{\mathrm{L}} \mid \psi \in Diff\mathbf{M} \} \in Riem^{\mathrm{L}}\mathbf{M}/Diff\mathbf{M} \ , \tag{6.2}$$

of Lorentzian metrics in $\mathbf{M}$, which constitutes a gauge orbit $(\mathbf{M}, \mathcal{g}^{\mathrm{L}})$ within the principal bundle $Diff\mathbf{M} \to Riem^{\mathrm{L}}\mathbf{M} \to Riem^{\mathrm{L}}\mathbf{M}/Diff\mathbf{M}$.

The extrapolation of this fact to quantum gravity can be achieved by considering that such a CGR gauge orbit provides a *metrization* in $\mathbf{M}$. This metrization is mathematically represented by the reduction of the general linear frame bundle $GL\mathbf{M}$ to the family of all Lorentz frame bundles $L\mathbf{M}(\boldsymbol{g}^{\mathrm{L}})$ with $\boldsymbol{g}^{\mathrm{L}} \in \mathcal{g}^{\mathrm{L}}$. Physically, such a metrization is tantamount to an operational verification aimed at establishing which *physical* representatives of linear frames $\{\boldsymbol{e}_0(x),\dots,\boldsymbol{e}_3(x)\}$ in $GL\mathbf{M}$ are actually local Lorentz frames in the *physically existing* metric. Indeed, given the fact that the *existence* of such a metric is a feature of the physical reality around us, rather than a matter of mere choice left to the discretion of the experimenter, or of the observer of a given physical occurrence, from the point of view of CGR such a procedure amounts to verifying which ones of the linear *macro*-frames, consisting of “rigid rods” and “standard clocks,” are Lorentzian in the operational sense originally stipulated by Einstein. However, if a relabeling of all frames, test bodies and fields is carried out, this *physical* verification cannot distinguish between all the various choices of *mathematical* labels $\{\boldsymbol{e}_0(x),\dots\boldsymbol{e}_3(x)\}$, if such relabelings are related by means of diffeomorphisms that give rise to various equivalent Lorentzian metric representatives. Moreover, even in those cases where the mathematical labeling of frames, test bodies and fields is kept fixed, by the strong equivalence principle *physically* equivalent descriptions of natural phenomena are still obtained for physically distinct choices of inertial Lorentz frames, if those frames in free fall are related by Lorentz gauge transformations.

The adaptation of these basic facts to quantum *micro*-frames leads,[9,10] in broad outline, to the following formulation of *quantum* geometrodynamic evolution: the presence of a specific state of the quantum gravitational field $\mathbf{g}^{\mathrm{Q}}$ within a base segment $\mathbf{S}_n$ dictates the quantum-geometric evolution of all the quantum “matter” fields from the inflow hypersurface $\Sigma_{t_{n-1}}$ to the outflow hypersurface $\Sigma_{t_n}$ of that base-segment; the values of the local quantum energy-momentum density operators of all the quantum “matter” fields in the resulting states of those fields along $\Sigma_{t_n}$ then *creates* a QGR gauge orbit $(\mathbf{S}_{n+1}, \mathcal{g}^{\mathrm{M}})$ by metrizing the quantum frame bundles over the subsequent base segment $\mathbf{S}_{n+1}$; this determines the formation of local states of the quantum gravitational field $\mathbf{g}^{\mathrm{Q}}$ within that base-segment (whose mean values provide the Lorentzian metrics in $\mathcal{g}^{\mathrm{M}}$), and singles out within the quantum gravitational fibres above its points $x$ the multi-graviton states of quantum gravitational radiation. This process then repeats itself from segment to segment, with the limit to “infinitely-thin” segments being eventually taken.

Of course, the presence of constraints and quantum gravitational self-interaction has to be taken into account before this very sketchy picture of quantum-geometric gravitational evolution can be completed into a legitimate extrapolation to the quantum regime of Einstein's formulation of classical gravity. This involves the extension of the base-segments $\mathbf{S}_n$ into supermanifolds $\boldsymbol{S}_n$ over which a quantum gravitational superbundle can be defined, and on whose sections a quantum gravitational gauge supergroup can act. As we shall see in Sec. 9, this quantum gravitational supergroup incorporates diffeomorphism invariance as well as Poincaré gauge invariance.

The resulting framework shares common features with both covariant and canonical quantum gravity, but it is also fundamentally distinct in many respects. Thus, in geometric quantum gravity equivalence classes $\mathcal{g}^{\mathrm{M}}$ of mean metrics are generated by the quantum geometrodynamic evolution of gravitational fields in mutual interaction "matter" fields, so that no "background" metric is prescribed – as is the case in covariant quantum gravity. Moreover, the presence of *quantum* (super)-frames also resolves the "issue of time" – which is much-debated in contemporary canonical gravity.[29] This solution emerges from de Broglie's fundamental idea[30] that, on account of its rest mass $m$, each massive elementary quantum object represents a natural clock with period $T = 2\pi/m$ in Planck natural units. The replacement of global with local frames insures that such a proper time is kept *locally*, in accordance with the fundamental features of Einstein's *general* relativistic conceptualization of an ultimately valid description of nature. On the other hand, the use of bundles of quantum frames or superframes which incorporate the fundamental length $\ell$ into their very structure eliminates all need for de Broglie's "pilot waves." Thus, the replacement of classical geometries, upon which de Broglie founded his ideas, with quantum geometries, obviates the need for extraneous assumptions about *quantum* reality in the same manner in which the introduction of the geometry of Minkowski space obviated the need for regarding Lorentz contractions of (classically visualized) electrons as real phenomena.

In the subsequent sections we shall outline how these fundamental physical ideas can be described in a mathematically precise manner (cf. Chapter 8 of Ref. 10 for the details).

## 7. Graviton Fibres and Their Internal Gauges

We initiate the geometro-stochastic quantization of gravity by considering a generic QGR gauge orbit $(\mathbf{S}, \mathcal{g}^{\mathrm{M}})$, generated during the course of geometrodynamic evolution. The base-segment $\mathbf{S}$ in such a gauge orbit is a differential manifold, whose boundary $\partial\mathbf{S}$ is composed of two disjoint hypersurfaces $\Sigma'$ and $\Sigma''$ that constitute its inflow and outflow hypersurfaces for quantum data emerging in the course of that evolution, and of an equivalence class $\mathcal{g}^{\mathrm{M}}$ of mean metrics, which gives rise to the metrized bundle $P^*\mathbf{S}^{\uparrow}(\mathcal{g}^{\mathrm{M}})$ of (affine) Poincaré coframes $\boldsymbol{u} = \{\boldsymbol{a}^*(x), \theta^i(x)\}$ above $\mathbf{S}$.

We can associate with each such coframe $\boldsymbol{u} \in P^*\mathbf{S}^{\uparrow}(\mathcal{g}^{\mathrm{M}})$ a single-graviton fibre $\mathbf{Z}_{\boldsymbol{u}}$ consisting of the graviton state vectors represented by the wave functions

$$\boldsymbol{f}(\boldsymbol{u};\zeta) = (2\pi)^{-3/2} \int_{\boldsymbol{V}_0^+} \exp\left(-i\,\zeta \cdot k\right) \tilde{\boldsymbol{f}}(\boldsymbol{u};k)\; d\Omega_0(k) \;, \qquad \zeta = q - i\ell v \;, \tag{7.1}$$

that are generated by all the momentum-space wave functions

$$\tilde{\boldsymbol{f}}(\boldsymbol{u};k) = \tilde{f}_{ij}(k)\, \boldsymbol{\theta}^i(x) \otimes \boldsymbol{\theta}^j(x) \;, \qquad \tilde{f}_{ij}(k) \equiv \tilde{f}_{ji}(k) \;, \tag{7.2}$$

in the mass-0 and spin-2 pseudo-unitary representation

$$\tilde{U}(a,\Lambda) \;:\; \tilde{f}_{ij}(k) \;\mapsto\; \tilde{f}'_{ij}(k) = \exp(ia \cdot k)\, \Lambda_i^{\;i'} \Lambda_j^{\;j'} \tilde{f}_{i'j'}(\Lambda^{-1}k) \tag{7.3}$$

of the orthochronous Poincaré group $\mathrm{ISO}^{\uparrow}(3,1)$. Each such fibre $\mathbf{Z}_{\boldsymbol{u}}$ is a Krein space,[31] i.e., it is a pseudo-Hilbert space, which in the present context carries the indefinite inner product

$$\langle \boldsymbol{f} | \boldsymbol{f}' \rangle = \int_{\boldsymbol{V}_0^+} \eta^{ii'} \eta^{jj'} \tilde{f}^*_{ij}(k)\, \tilde{f}'_{i'j'}(k)\, d\Omega_0(k) \;, \quad d\Omega_0(k) = \delta(k^2)\, d^4k \;, \tag{7.4}$$

and the positive-definite $J_{\boldsymbol{u}}$-inner product associated with the coframe $\boldsymbol{u} \in P^*\mathbf{S}^{\uparrow}(\mathbf{g}^{\mathrm{M}})$:

$$\big(\boldsymbol{f}|\boldsymbol{f}'\big)_{J_{\boldsymbol{u}}} = \sum_{i,j=0}^{3} \int_{\mathbf{V}_0^+} \tilde{f}_{ij}^{*}(k)\, \tilde{f}'_{ij}(k)\, d\Omega_0(k) \ . \tag{7.5}$$

The former is left invariant by the representation resulting from (7.3),

$$U(a,\Lambda)\ :\ f_{ij}(q-i\ell v)\ \mapsto\ f'_{ij}(q-i\ell v) = \Lambda_i^{\ i'}\Lambda_j^{\ j'} f_{i'j'}(\Lambda^{-1}(q-a)-i\ell\Lambda^{-1}v)\ ; \tag{7.6}$$

whereas, the latter is not invariant under Lorentz boosts, but the majorant topology[31] to which that $\boldsymbol{J_u}$-inner product gives rise is left invariant by this representation.

Let us now introduce the generalized Lorenz gauge subfibres of $\mathbf{Z}_{\boldsymbol{u}}$ ,

$$\mathbf{Z}_{\boldsymbol{u}}^{(b)} = \left\{ \boldsymbol{f} \in \mathbf{Z}_{\boldsymbol{u}} \,\middle|\, \partial^i f_{ij} = b_j \ \ , \ \ j = 0,1,2,3 \right\} \ , \qquad \partial^i = \partial/\partial q^i . \tag{7.7}$$

They are obviously left invariant by the internal Lorenz gauge transformations

$$\boldsymbol{f} = f_{ij}\, \boldsymbol{\theta}^i(x) \otimes \boldsymbol{\theta}^j(x) \ \ \mapsto\ \ \boldsymbol{f}' = f'_{ij}\, \boldsymbol{\theta}^i(x) \otimes \boldsymbol{\theta}^j(x) \ , \qquad \boldsymbol{f}, \boldsymbol{f}' \in \mathbf{Z}_{\boldsymbol{u}} \ , \tag{7.8}$$

$$f'_{ij}(\zeta) = f_{ij}(\zeta) + \partial_i \lambda_j(\zeta) + \partial_j \lambda_i(\zeta)\ \ , \qquad \partial^i \partial_i \lambda_j(\zeta) = 0\ \ , \qquad \partial^i \lambda_i(\zeta) = 0\ . \tag{7.9}$$

The generic element $\boldsymbol{f}$ of $\mathbf{Z}_{\boldsymbol{u}}$ can be decomposed in the following unique manner,

$$\boldsymbol{f} = \boldsymbol{f}^{\mathrm{TT}} + \boldsymbol{f}^{\perp} + \boldsymbol{f}^{b} \in \mathbf{Z}_{\boldsymbol{u}}^{(b)}\ \ , \qquad \boldsymbol{f}^{\mathrm{TT}} \in \mathbf{Z}_{\boldsymbol{u}}^{\mathrm{TT}}\ , \qquad \boldsymbol{f}^{\perp} \in \mathbf{N}_{\boldsymbol{u}}\ , \tag{7.10}$$

where $\boldsymbol{f}^{\mathrm{TT}}$ is the element of the TT-subfibre of $\mathbf{Z}_{\boldsymbol{u}}$, whose elements satisfy the transverse and traceless conditions that are well-known from linearized gravity, and can be expressed as a superposition of two *physical* (linear or circular) modes of polarization; whereas, the pseudo-orthogonal complement of $\boldsymbol{f}^{\mathrm{TT}}$ belongs to the null-subfibre

$$\mathbf{N}_{\boldsymbol{u}} = \left\{ \boldsymbol{f} \in \mathbf{Z}_{\boldsymbol{u}} \,\middle|\, \langle \boldsymbol{f} | \boldsymbol{f} \rangle = 0 \right\} \tag{7.11}$$

of $\mathbf{Z}_{\boldsymbol{u}}$, which can be shown to be a subspace of the proper Lorenz gauge subfibre corresponding to $b_j(\zeta) \equiv 0, j = 0,1,2,3,$ in Eq. (7.7).

Let us introduce the following expansion of a superlocal (i.e., local in $P^{*}\mathbf{S}^{\uparrow}(\mathbf{g}^{\mathrm{M}})$, and not just in $\mathbf{S}$) graviton wave function,

$$\boldsymbol{f}(\boldsymbol{u};\zeta) = (2\pi)^{-3/2} \int_{\mathbf{V}_0^+} \exp\!\big(-i\,\zeta \cdot k\big)\, \boldsymbol{\varepsilon}^{[r]}(\boldsymbol{u};k) \otimes \boldsymbol{\varepsilon}^{[s]}(\boldsymbol{u};k)\, \tilde{f}_{[rs]}(k)\, d\Omega_0(k)\ \ , \tag{7.12}$$

with respect to the graviton polarization null-frame whose elements are given by

$$\boldsymbol{\varepsilon}^{[rs]\zeta}(\boldsymbol{u};\zeta') = \int_{\mathbf{V}_0^+} \exp[i(\bar{\zeta} - \zeta') \cdot k]\ \boldsymbol{\varepsilon}^{[r]}(\boldsymbol{u};k) \otimes \boldsymbol{\varepsilon}^{[s]}(\boldsymbol{u};k)\, d\Omega_0(k)\ , \quad r,s = \pm 1, \pm 2. \tag{7.13}$$

This graviton polarization null-frame is obtained from the null polarization cotetrad whose elements within the complexified cotangent space $\overline{T}_x{}^{*}\mathbf{S}$,

$$\boldsymbol{\varepsilon}^{[\pm 1]}(\boldsymbol{u};k) = \mp \tfrac{1}{\sqrt{2}} \Big( \boldsymbol{\varepsilon}^{(1)}(\boldsymbol{u};k) \pm i \boldsymbol{\varepsilon}^{(2)}(\boldsymbol{u};k) \Big) = \mp \tfrac{1}{\sqrt{2}} \Big( \boldsymbol{\theta}^1(x) \pm i\, \boldsymbol{\theta}^2(x) \Big) \cdot R(\hat{\mathbf{k}})\ , \tag{7.14}$$

$$\boldsymbol{\varepsilon}^{[+2]}(\boldsymbol{u};k) = k_i\, \boldsymbol{\theta}^i(x) = \boldsymbol{k}\ \ , \qquad \boldsymbol{\varepsilon}^{[-2]}(\boldsymbol{u};k) = k_0^{-1} \boldsymbol{\theta}^0(x) - \tfrac{1}{2} k_0^{-2}\, \boldsymbol{k}\ \ , \tag{7.15}$$

are the outcome[9,10] of a standard application of Wigner rotations to the linear coframe $\{\theta^i(x)\}$ within the affine coframe $\boldsymbol{u}$. Then the TT-gauge subfibre appearing in Eq. (7.10) can be expressed as follows:[10]

$$\mathbf{Z}_{\boldsymbol{u}}^{\mathrm{TT}} = \left\{ \sum_{r=-1}^{+1} f_{[rr]\zeta}\, \boldsymbol{\varepsilon}^{[rr]\zeta} \,\middle|\, \boldsymbol{f} = f_{[rs]\zeta}\, \boldsymbol{\varepsilon}^{[rs]\zeta} \in \mathbf{Z}_{\boldsymbol{u}}^{(0)} \right\} . \tag{7.16}$$

The proper Lorenz gauge subfibre of $\mathbf{Z}_{\boldsymbol{u}}$ that contains the TT-gauge subfibre in Eq. (7.16) is left invariant by the representation in (7.6). In case that $(a,\Lambda)$ in Eq. (7.6) is allowed to vary only over the proper Poincaré group $\mathrm{ISO}_0(3,1)$, the restriction of the resulting representation of $\mathrm{ISO}_0(3,1)$ to the proper Lorenz gauge subfibre of $\mathbf{Z}_{\boldsymbol{u}}$ can be shown[10] to be equivalent the Wigner-type representation of that group given by the following transformations under the action of $(a,\Lambda) \in \mathrm{ISO}_0(3,1)$,

$$\alpha_\pm(q - i\ell v) \;\mapsto\; e^{\pm 2i\theta(\Lambda, i\partial)}\,\alpha_\pm(\Lambda^{-1}(q-a) - i\ell\Lambda^{-1}v)\ , \qquad \partial := (\partial_0, \partial_1, \partial_2, \partial_3)\ . \tag{7.17}$$

The wave functions in Eq (7.17) are the two graviton circular polarization components

$$\alpha_\pm(\zeta) = f_{[\pm1\pm1]\zeta}\ , \qquad f_{[rs]\zeta} = \eta_{rr'}\eta_{ss'}\left\langle \boldsymbol{\varepsilon}^{[r's']\zeta} \middle| \boldsymbol{f} \right\rangle, \quad r,s = \pm1, \pm2\ , \tag{7.18}$$

contained in the expansion in Eq. (7.12) of a graviton state vector *above* $\boldsymbol{u} \in P^*\mathbf{S}^{\uparrow}(\mathcal{g}^{\mathrm{M}})$.

## 8. Quantum Gravitational and Berezin-Faddeev-Popov Frames

To carry out the geometro-stochastic second-quantization of gravity, we construct a quantum gravitational fibre $\mathcal{Z}_{\boldsymbol{u}}$ above each Poincaré coframe $\boldsymbol{u} = \{\boldsymbol{a}^*(x), \theta^i(x)\} \in P^*\mathbf{S}^{\uparrow}(\mathbf{g}^{\mathrm{M}})$ by assigning to each $\boldsymbol{u}$ a superlocal vacuum state vector $\Psi_{0;\boldsymbol{u}}$, that spans the one-dimensional superlocal vacuum sector $\mathcal{Z}_{0;\boldsymbol{u}}$. We then use the $\boldsymbol{u}$-dependent $J_{\boldsymbol{u}}$-inner product in Eq. (7.5) to add to $\mathcal{Z}_{0;\boldsymbol{u}}$ the $J_{\boldsymbol{u}}$-direct sums of symmetrized tensor products of the graviton fibre $\mathbf{Z}_{\boldsymbol{u}}$, thus arriving at the quantum gravitational fibre

$$\mathcal{Z}_{\boldsymbol{u}} = \bigoplus_{J}{}_{n=0}^{\infty}\ \mathcal{Z}_{n;\boldsymbol{u}}\ , \qquad \mathcal{Z}_{n;\boldsymbol{u}} = \mathbf{Z}_{\boldsymbol{u}} \underset{S}{\otimes} \cdots \underset{S}{\otimes} \mathbf{Z}_{\boldsymbol{u}}\ . \tag{8.1}$$

Within this quantum gravitational fibre we can define graviton annihilation operators that act on $n$-graviton state vectors belonging to $\mathcal{Z}_{n;\boldsymbol{u}}$ as follows,

$$\left(g_{ij}^{(-)}(\boldsymbol{u};\zeta)\,\Psi_{n;\boldsymbol{u}}\right)_{n-1}(\zeta_1, i_1j_1, \ldots, \zeta_{n-1}, i_{n-1}j_{n-1}) = \sqrt{n}\ \Psi_{n;\boldsymbol{u}}(\zeta, ij, \zeta_1, i_1j_1, \ldots, \zeta_{n-1}, i_{n-1}j_{n-1}), \tag{8.2}$$

and corresponding graviton creation operators which are their $J_{\boldsymbol{u}}$-adjoints.[31] We can then introduce in Lorenz quantum gravitational subfibre of $\mathcal{Z}_{\boldsymbol{u}}$ the elements

$$\boldsymbol{\Phi}_{\boldsymbol{f}} = \exp\left[-\tfrac{1}{2}\langle \boldsymbol{f}|\boldsymbol{f}\rangle + g^{(+)}(\boldsymbol{u};\boldsymbol{f})\right]\ \boldsymbol{\Psi}_{0;\boldsymbol{u}}\ , \qquad \boldsymbol{f} \in \mathbf{Z}_{\boldsymbol{u}}^{(0)}\ , \tag{8.3}$$

of a quantum gravitational frame, defined by the general procedure used in Eq. (4.13). However, due to the presence of an indefinite metric, this quantum gravitational frame provides the continuous resolution of the identity

$$\int_{\mathbf{Z}_{\boldsymbol{u}}^{\mathrm{TT}}} \left|\boldsymbol{\Phi}_{\boldsymbol{f}}\right\rangle\ d\boldsymbol{f}\, d\bar{\boldsymbol{f}} \left\langle \boldsymbol{\Phi}_{\boldsymbol{f}}\right| \ = \ \mathbf{1}_{\boldsymbol{u}}^{\mathrm{TT}}, \tag{8.4}$$

only within the *physical* quantum gravitational fibre above $\boldsymbol{u}$, corresponding to the following $J_{\boldsymbol{u}}$-orthogonal decomposition of the Lorenz quantum gravitational subfibre:

$$\mathcal{Z}_{\boldsymbol{u}}^{(0)} = \mathcal{Z}_{\boldsymbol{u}}^{\mathrm{TT}} \underset{J}{\oplus} \mathcal{N}_{\boldsymbol{u}}\,, \qquad \mathcal{N}_{\boldsymbol{u}} = \left\{\ \boldsymbol{\Psi} \in \mathcal{Z}_{\boldsymbol{u}}^{(0)} \,\middle|\, \langle \boldsymbol{\Psi}|\boldsymbol{\Psi}\rangle = 0\right\}\ . \tag{8.5}$$

To treat the remaining degrees of freedom of graviton polarization, representing *un*physical modes, we shall use adaptations[9,10] to the quantum-geometric framework of quantum nonabelian gauge-field theoretical techniques.[32] Thus, we shall introduce above each $\boldsymbol{u} \in P^*\mathbf{S}^{\uparrow}(\mathbf{g}^{\mathrm{M}})$ the graviton ghost and antighost fibres

$$\mathbf{F}_{\boldsymbol{u}} = \left\{\mathbf{c} = c_i\,\boldsymbol{\theta}^i(x)\ \middle|\ (c_0, \ldots, c_3) \in \mathbf{K}\right\}, \quad \overline{\mathbf{F}}_{\boldsymbol{u}} = \left\{\bar{\mathbf{c}} = \bar{c}_i\,\boldsymbol{\theta}^i(x) \middle|\ (\bar{c}_0, \ldots, \bar{c}_3) \in \mathbf{K}^{\dagger}\right\}\ , \tag{8.6}$$

where $\mathbf{K}$ is a Krein space (with $\mathbf{K}^{\dagger}$ denoting its dual) consisting of functions in terms of which we can carry out for the decompositions in Eq. (7.10) the identifications

$$\boldsymbol{f}^{\perp} = f_{ij}^{\perp}\, \boldsymbol{\theta}^i(x) \otimes \boldsymbol{\theta}^j(x) \;\leftrightarrow\; \mathbf{c}^{\perp} \in \mathbf{F}_{\boldsymbol{u}} \;, \quad f_{ij}^{\perp} = \partial_i c_j^{\perp} + \partial_j c_i^{\perp} \;, \quad \partial^i c_i^{\perp} = 0 \;, \tag{8.7}$$

$$\boldsymbol{f}^{b} = f_{ij}^{b}\, \boldsymbol{\theta}^i(x) \otimes \boldsymbol{\theta}^j(x) \;\leftrightarrow\; \mathbf{c}^{b} \in \mathbf{F}_{\boldsymbol{u}} \;, \quad f_{ij}^{b} = \partial_i c_j^{b} + \partial_j c_i^{b} \;, \quad \partial_j \partial^i c_i^{b} = b_j \;. \tag{8.8}$$

We can then construct Faddeev-Popov (FP) ghost and antighost fibres,

$$\mathcal{F}_{\boldsymbol{u}} = \bigoplus\nolimits_{n=0}^{\infty} (\mathbf{F}_{\boldsymbol{u}} \underset{A}{\otimes} \cdots \underset{A}{\otimes} \mathbf{F}_{\boldsymbol{u}})_n \;, \qquad \overline{\mathcal{F}}_{\boldsymbol{u}} = \bigoplus\nolimits_{n=0}^{\infty} (\overline{\mathbf{F}}_{\boldsymbol{u}} \underset{A}{\otimes} \cdots \underset{A}{\otimes} \overline{\mathbf{F}}_{\boldsymbol{u}})_n \;, \tag{8.9}$$

in which we can define in the standard manner the fermionic-type action of the ghost annihilation and creation operators $C^{(-)}(\boldsymbol{u};\zeta)$ and $C^{(+)}(\boldsymbol{u};\zeta)$, respectively, together with that of the conjugate operators $\overline{C}^{(-)}(\boldsymbol{u};\zeta)$ and $\overline{C}^{(+)}(\boldsymbol{u};\zeta)$ acting on antighost states.

By following procedures that emerge from the standard techniques that are employed in Berezin integration,[25] we make the transition to the BFP (i.e., Berezin-Faddeev-Popov) ghost-antighost superfibres $\mathcal{B}_{\boldsymbol{u}} \otimes \overline{\mathcal{B}}_{\boldsymbol{u}}$ obtained by introducing a $J_{\boldsymbol{u}}$-orthonormal basis $\{\mathbf{c}_{\boldsymbol{u};1}, \mathbf{c}_{\boldsymbol{u};2}, \ldots\}$ within each $\mathbf{F}_{\boldsymbol{u}}$, and its conjugate within $\overline{\mathbf{F}}_{\boldsymbol{u}}$, and then setting

$$\mathbf{B}_{\boldsymbol{u}} = \left\{ \boldsymbol{c} = \theta_{\boldsymbol{u}}^{\alpha}\, \mathbf{c}_{\boldsymbol{u};\alpha} \,\middle|\, \theta_{\boldsymbol{u}}^{\alpha} = \left(\mathbf{c}_{\boldsymbol{u};\alpha} | \mathbf{c}\right)_{J_{\boldsymbol{u}}} \vartheta_{\alpha} \;, \;\; \alpha = 1,2,\ldots \;, \;\; \mathbf{c} \in \mathbf{F}_{\boldsymbol{u}} \right\} , \tag{8.10}$$

$$\overline{\mathbf{B}}_{\boldsymbol{u}} = \left\{ \bar{\boldsymbol{c}} = \bar{\mathbf{c}}_{\boldsymbol{u};\alpha} \bar{\theta}_{\boldsymbol{u}}^{\alpha} \,\middle|\, \bar{\theta}_{\boldsymbol{u}}^{\alpha} = \left(\bar{\mathbf{c}}_{\boldsymbol{u};\alpha} | \bar{\mathbf{c}}\right)_{J_{\boldsymbol{u}}} \bar{\vartheta}_{\alpha} \;, \;\; \alpha = 1,2,\ldots \;, \;\; \bar{\mathbf{c}} \in \overline{\mathbf{F}}_{\boldsymbol{u}} \right\} , \tag{8.11}$$

where $\vartheta_{\alpha}$ and $\bar{\vartheta}_{\alpha}$ are the generators of a Grassmann algebra. Within these superfibres we introduce the ghost and antighost BFP superframe elements

$$\boldsymbol{\Phi}_{\boldsymbol{c}} = \exp\left(-\tfrac{1}{2}\sum\nolimits_{\alpha=1}^{\infty} \bar{\theta}_{\boldsymbol{u}}^{\alpha}\, \theta_{\boldsymbol{u}}^{\alpha}\right) \exp\left(C^{(+)}(\boldsymbol{u};\boldsymbol{c})\right) \boldsymbol{\Psi}_{0;\boldsymbol{u}} \;, \quad \boldsymbol{c} \in \mathbf{B}_{\boldsymbol{u}} \;, \tag{8.12}$$

$$\boldsymbol{\Phi}_{\bar{\boldsymbol{c}}} = \exp\left(-\tfrac{1}{2}\sum\nolimits_{\alpha=1}^{\infty} \theta_{\boldsymbol{u}}^{\alpha}\, \bar{\theta}_{\boldsymbol{u}}^{\alpha}\right) \exp\left(\overline{C}^{(+)}(\boldsymbol{u};\bar{\boldsymbol{c}})\right) \boldsymbol{\Psi}_{0;\boldsymbol{u}} \;, \quad \bar{\boldsymbol{c}} \in \overline{\mathbf{B}}_{\boldsymbol{u}} \;. \tag{8.13}$$

The resulting superframes supply continuous resolutions of the identity within the corresponding FP ghost and antighost fibres in Eq. (8.9):

$$\int_{\overline{\mathbf{B}}_{\boldsymbol{u}} \times \mathbf{B}_{\boldsymbol{u}}} \langle \boldsymbol{\Psi} | \boldsymbol{\Phi}_{\boldsymbol{c}} \rangle\, \boldsymbol{d\bar{c}}\, \boldsymbol{dc} \langle \boldsymbol{\Phi}_{\boldsymbol{c}} | \boldsymbol{\Psi}' \rangle = \langle \boldsymbol{\Psi} | \boldsymbol{\Psi}' \rangle \;, \qquad \boldsymbol{\Psi}, \boldsymbol{\Psi}' \in \mathcal{F}_{\boldsymbol{u}} \;, \tag{8.14}$$

$$\int_{\mathbf{B}_{\boldsymbol{u}} \times \overline{\mathbf{B}}_{\boldsymbol{u}}} \langle \boldsymbol{\Psi} | \boldsymbol{\Phi}_{\bar{\boldsymbol{c}}} \rangle\, \boldsymbol{dc}\, \boldsymbol{d\bar{c}} \langle \boldsymbol{\Phi}_{\bar{\boldsymbol{c}}} | \boldsymbol{\Psi}' \rangle = \langle \boldsymbol{\Psi} | \boldsymbol{\Psi}' \rangle \;, \qquad \boldsymbol{\Psi}, \boldsymbol{\Psi}' \in \overline{\mathcal{F}}_{\boldsymbol{u}} \;. \tag{8.15}$$

The fundamental significance of the quantum gravitational frames and of the BFP superframes introduced in this section to the formulation in Sec. 11 of quantum geometrodynamic evolution is that, in addition to supplying continuous resolutions of the identity, the elements of these frames and superframes are also eigenvectors of corresponding annihilation operators, so that we have:

$$g_{ij}^{(-)}(\boldsymbol{u};\zeta)\, \boldsymbol{\Phi}_{\boldsymbol{f}} = f_{ij}(\zeta)\, \boldsymbol{\Phi}_{\boldsymbol{f}} \;, \qquad \boldsymbol{f} = f_{ij}\, \boldsymbol{\theta}^i(x) \otimes \boldsymbol{\theta}^j(x) \in \mathbf{Z}_{\boldsymbol{u}}^{(0)} \;, \tag{8.16}$$

$$C_i^{(-)}(\boldsymbol{u};\zeta)\boldsymbol{\Phi}_{\boldsymbol{c}} = c(\zeta,i)\boldsymbol{\Phi}_{\boldsymbol{c}} \;, \qquad \overline{C}_i^{(-)}(\boldsymbol{u};\zeta)\boldsymbol{\Phi}_{\bar{\boldsymbol{c}}} = \bar{c}(\zeta,i)\boldsymbol{\Phi}_{\bar{\boldsymbol{c}}} \;. \tag{8.17}$$

These features mediate the extension of the techniques of Sec. 5 to quantum gravity.

## 9. The Quantum Gravitational Supergroup

Each of the TT-gauge subfibres of $\mathbf{Z}_{\boldsymbol{u}}$, defined in Eq. (7.16), is left invariant by the semi-direct product of the spacetime translation group $T^4$ and the rotation group O(3) viewed as a subgroup of the orthochronous Poincaré group $ISO^{\uparrow}(3,1)$, but not by the Lorentz boosts $\Lambda_v$ within $ISO^{\uparrow}(3,1)$. Hence, we shall carry out the decompositions

$$(a,\Lambda) = (a,R)(0,\Lambda_v) \in ISO^{\uparrow}(3,1)\ ,\ \ (a,R) \in T^4 \wedge O(3)\ ,\ \ \Lambda_v \in SO_0(3,1)\ , \quad (9.1)$$

$$(a,\Lambda)^{-1} = (0,\Lambda_v^{-1})(-R^{-1}a,R^{-1}) \in ISO^{\uparrow}(3,1)\ ,\ \ (-R^{-1}a,R^{-1}) = (a,R)^{-1}\ , \quad (9.2)$$

of a generic orthochronous Poincaré transformation and its inverse. Then, if we introduce the quantum gravitational superfibres

$$\mathcal{Z}_{\boldsymbol{u}}^{\Diamond} = \mathcal{Z}_{\boldsymbol{u}}^{TT} \otimes \overline{\mathcal{B}}_{\boldsymbol{u}} \otimes \mathcal{B}_{\boldsymbol{u}}\ , \quad (9.3)$$

and consider within them the *quantum gravitational coframes* with elements

$$\boldsymbol{\Phi}^{\dagger}_{\boldsymbol{f},\bar{\mathbf{c}},\mathbf{b},\mathbf{c}} = \boldsymbol{\Phi}^{\dagger}_{\boldsymbol{f}} \otimes \boldsymbol{\Phi}_{\bar{\mathbf{c}}} \otimes \boldsymbol{\Phi}^{\dagger}_{\mathbf{b},\mathbf{c}}\ ,\quad \boldsymbol{f} \in \mathbf{Z}_{\boldsymbol{u}}^{TT},\quad \mathbf{b} \in \mathbf{F}_{\boldsymbol{u}}\ ,\quad \bar{\mathbf{c}} \in \overline{\mathbf{F}}_{\boldsymbol{u}}^{\perp}\ ,\quad \mathbf{c} \in \mathbf{F}_{\boldsymbol{u}}^{\perp}\ , \quad (9.4)$$

we can take advantage of Eqs. (9.1) and (9.2) to transfer the action of a Lorentz boost $\Lambda_v$ upon a (classically conceptualized) coframe $\boldsymbol{u} \in P^*\mathbf{S}^{\uparrow}(\mathbf{g}^{M})$ into the action upon the corresponding quantum gravitational coframe in Eq. (9.4), carried out via the actions

$$U(0,\Lambda_v) : \boldsymbol{c}^{\perp} = \theta_{\boldsymbol{u}}^{\alpha}\mathbf{c}_{\boldsymbol{u};\alpha} \mapsto \theta'^{\alpha}_{\boldsymbol{u}}\mathbf{c}_{\boldsymbol{u};\alpha}\ ,\quad \theta'^{\alpha}_{\boldsymbol{u}} = \left(\mathbf{c}_{\boldsymbol{u};\alpha}\middle|\boldsymbol{c}^{\perp}\right)_{J_{\boldsymbol{u}}} U^{\beta}_{\alpha}(0,\Lambda_v)\vartheta_{\beta}\ , \quad (9.5)$$

$$U(0,\Lambda_v) : \boldsymbol{c}^{b} = \theta_{\boldsymbol{u}}^{\alpha}\mathbf{c}_{\boldsymbol{u};\alpha} \mapsto \theta'^{\alpha}_{\boldsymbol{u}}\mathbf{c}_{\boldsymbol{u};\alpha}\ ,\quad \theta'^{\alpha}_{\boldsymbol{u}} = \left(\mathbf{c}_{\boldsymbol{u};\alpha}\middle|\mathbf{c}^{b}\right)_{J_{\boldsymbol{u}}} U^{\beta}_{\alpha}(0,\Lambda_v)\vartheta_{\beta}\ , \quad (9.6)$$

$$U^{\alpha}_{\beta}(0,\Lambda_v) = \left(\mathbf{c}_{\boldsymbol{u};\beta}\middle|U(0,\Lambda_v)\mathbf{c}_{\boldsymbol{u};\alpha}\right)_{J_{\boldsymbol{u}}}\ ,\qquad \alpha,\beta = 1,2,\dots\ , \quad (9.7)$$

upon its superlocal ghost states (associated, in accordance with Eqs. (8.7) and (8.8), with the decomposition in Eq. (7.10) in any given generalized internal Lorenz gauge), and to corresponding actions upon its superlocal antighost states. This constitutes the first step in the construction of a quantum gravitational supergroup that represents the counterpart of the CGR gauge group given by the semidirect product

$$\mathcal{G}(\mathbf{M},\mathcal{g}^{L}) = G_A(\mathit{Diff}\,\mathbf{M}) \wedge \mathcal{G}(\mathbf{P}(\mathbf{M},ISO^{\uparrow}(3,1))\ , \quad (9.8)$$

which corresponds to the equivalence class $\mathcal{g}^{L}$ of physically equivalent Lorentzian metrics of a CGR model over a classical spacetime manifold $\mathbf{M}$ (cf. Sec. 6).

To understand the physical significance of the CGR gauge group in Eq. (9.8), consider the general affine coframe bundle $GA^*\mathbf{M}$ consisting of all the affine coframes $\{\boldsymbol{a}^*(x),\theta^i(x)\}$, $x \in \mathbf{M}$, over the manifold $\mathbf{M}$. Let us then express the notion of diffeomorphism invariance discussed in Sec. 6 in terms of the vertical automorphisms determined by the following action of the diffeomorphisms $\psi \in \mathit{Diff}\,\mathbf{M}$ upon any section $\boldsymbol{s}$ of $GA^*\mathbf{M}$, representing a *moving* coframe in the sense of Cartan:[33]

$$\mathbf{D}(\psi)\ :\ \boldsymbol{s} = \{(\boldsymbol{a}^*,\theta^i)\} \mapsto \boldsymbol{s}' = \{(\psi_*(\boldsymbol{a}^* \circ \psi^{-1}),\ \psi_*(\theta^i \circ \psi^{-1})\}\ . \quad (9.9)$$

This action gives rise to a moving coframe $\boldsymbol{s}'$ with the same domain as that of $\boldsymbol{s}$, so that, upon covering $GA^*\mathbf{M}$ with a bundle atlas of local trivializations associated with local sections of $GA^*\mathbf{M}$, we can assign to each $\psi \in \mathit{Diff}\,\mathbf{M}$ a vertical automorphism $\psi \in \mathrm{Aut}_{\mathbf{M}}\mathbf{P}$. In this manner we arrive at an isomorphism between $\mathit{Diff}\,\mathbf{M}$ and a subgroup $G_A(\mathit{Diff}\,\mathbf{M})$ of the group $\mathrm{Aut}_{\mathbf{M}}\mathbf{P}$ of all vertical automorphisms in the principal bundle $\mathbf{P} = GA^*\mathbf{M}$.

Consider now the metrized affine coframe bundle $P^*\mathbf{M}^{\uparrow}(\mathcal{g}^{L})$ corresponding to the family of all Lorentz frame bundles $L\mathbf{M}(\boldsymbol{g}^{L})$, $\boldsymbol{g}^{L} \in \mathcal{g}^{L}$, described in Sec. 6. Let us call a cross-section of $P^*\mathbf{M}^{\uparrow}(\mathcal{g}^{\mathrm{L}})$ a *physical* moving coframe in $\mathbf{M}$ if it coincides with a cross-section $\boldsymbol{s}$ of one of its subbundles $P^*\mathbf{M}^{\uparrow}(\boldsymbol{g}^{\mathrm{L}})$. We shall denote such a physical moving coframe by $(\boldsymbol{s},\boldsymbol{g}^{\mathrm{L}})$, so that the corresponding cross-section $\{\theta^i(x)\,|\,x\in\mathbf{M}\}$ of $L^*\mathbf{M}^{\uparrow}(\boldsymbol{g}^{\mathrm{L}})$ represents in CGR a vierbein. The following action of the gauge group $\mathrm{Aut}_{\mathbf{M}}P\mathbf{M}^{\uparrow}(\boldsymbol{g}^{\mathrm{L}})$, which is isomorphic to the group $\mathcal{G}(\mathbf{P}(\mathbf{M},\mathrm{ISO}^{\uparrow}(3,1))$ of Poincaré gauge transformations appearing in Eq. (9.8), and the subgroup $\mathrm{G}_A(\mathit{Diff}\,\mathbf{M})$ of $\mathrm{Aut}_{\mathbf{M}}GA^*\mathbf{M}$ upon any such physical moving coframe $(\boldsymbol{s},\boldsymbol{g}^{\mathrm{L}})$,

$$(\psi,\phi):\ (\boldsymbol{s},\boldsymbol{g}^{\mathrm{L}})\ \mapsto\ (\,\boldsymbol{s}\circ\phi\circ\psi\,,\psi_*\,\boldsymbol{g}^{\mathrm{L}})\,,\quad \psi\in\mathrm{G}_A(\mathit{Diff}\mathbf{M})\,,\ \ \phi\in\mathrm{Aut}_{\mathbf{M}}P\mathbf{M}^{\uparrow}(\boldsymbol{g}^{\mathrm{L}})\,, \tag{9.10}$$

clearly represents the mathematical embodiment of the kind of generic change of moving frame of reference in CGR under which all *physical* predictions of a CGR model remain invariant. By Eq. (9.10), this combination of Poincaré gauge invariance with the diffeomorphism invariance gives rise to the group multiplication law

$$(\psi,\phi)(\psi',\phi')=((\psi\circ\phi')\circ\psi',\phi\phi')\,,\ \ \psi\,,\psi'\in\mathrm{G}_A(\mathit{Diff}\mathbf{M}),\ \ \phi\,,\phi'\in\mathcal{G}(\mathbf{P})\,, \tag{9.11}$$

where $\mathbf{P}=\mathbf{P}(\mathbf{M},\mathrm{ISO}^{\uparrow}(3,1))$, thus determining the CGR gauge group in Eq. (9.8).

Due to the fundamentally indeterministic character of quantum theory, which is based on a theory of measurement in which the quantum states of matter fields can undergo reductions that can be only probabilistically predicted, it would be inconsistent to assume that the structure of spacetime can be deterministically predicted at the microscopic level, since any change in those states is bound to affect its quantum geometric structure. Hence, local indeterministic changes in the *mean* metrics $\mathbf{g}^{\mathrm{M}}$ of a quantum gravitational field can occur with the passage of proper time of local observers, so that, as opposed to the situation in CGR, there cannot be a *global* counterpart $(\mathbf{M},\mathcal{g}^{\mathrm{M}})$ of the CGR gauge orbit $(\mathbf{M},\mathcal{g}^{\mathrm{L}})$, and *global* diffeomorphism invariance loses its *raison d'être* in the quantum gravitational regime. This implies that the only form of causality by which quantum geometrodynamic evolution can abide is the one in which the time-ordering of local events is correlated to the time-ordering of the inflow and outflow hypersurfaces $\Sigma'$ and $\Sigma''$ of base-segments $\mathbf{S}$ *created* during that evolution. In other words, any mean metric $\mathbf{g}^{\mathrm{M}}$ to which the inflow of matter and radiation gives rise has to be such that the futures of all points on $\Sigma'$ that are classically causal in relation $\mathbf{g}^{\mathrm{M}}$ intersect $\Sigma''$, and the pasts of all points on $\Sigma''$ that are classically causal in relation to the same $\mathbf{g}^{\mathrm{M}}$ intersect $\Sigma'$.

This implies that the quantum metrization of a general affine coframe bundle $GA^*\mathbf{S}$ selects a metrized coframe bundle $P^*\mathbf{S}^{\uparrow}(\mathcal{g}^{\mathrm{M}})$ corresponding to a family $\mathcal{g}^{\mathrm{M}}$ of Lorentzian mean metrics $\mathbf{g}^{\mathrm{M}}$ that are diffeomorphically equivalent under the group $\mathit{Diff}_0\,\mathbf{S}$ of diffeomorphisms of a base-segment $\mathbf{S}$ that leave its inflow and outflow hypersurfaces invariant (rather than under the group $\mathit{Diff}\,\mathbf{S}$ of all diffeomorphisms of $\mathbf{S}$), thus giving rise a QGR gauge orbit $(\mathbf{S},\mathcal{g}^{\mathrm{M}})$. Each diffeomorphism $\psi\in\mathit{Diff}_0\mathbf{S}$ gives rise, in the same manner as in the above considered classical context, to a vertical automorphism $\psi\in\mathrm{G}_A(\mathit{Diff}_0\mathbf{S})$. In general, such a $\psi$ acts on physical moving coframes given by cross-sections $\boldsymbol{s}$ of the Poincaré coframe bundle $P^*\mathbf{S}^{\uparrow}(\mathbf{g}^{\mathrm{M}})$, by mapping them, in accordance with Eq. (9.9), into physical moving coframes given by cross-sections $\boldsymbol{s}'$ of some other Poincaré coframe bundle $P^*\mathbf{S}^{\uparrow}(\mathbf{g}'^{\mathrm{M}})$ with $\mathbf{g}'^{\mathrm{M}}\in(\mathbf{S},\mathcal{g}^{\mathrm{M}})$. The family $\mathrm{G}_A(\mathit{Diff}_0\mathbf{S})$ of all such vertical automorphisms constitutes a subgroup of the vertical automorphism group $\mathrm{Aut}_{\mathbf{S}}\mathbf{P}$ for $\mathbf{P}=GA^*\mathbf{S}$.

We can proceed as in Eqs. (9.9)-(9.10) to define the combined action of quantum-geometrically causal diffeomorphisms and Poincaré gauge transformations

$(\psi,\phi)$: $(\boldsymbol{s},\mathbf{g}^{\mathrm{M}}) \mapsto (\boldsymbol{s}',\mathbf{g}'^{\mathrm{M}})$, $(\psi,\phi) \in \mathcal{G}(\mathbf{S},\mathcal{g}^{\mathrm{M}}) = \mathrm{G}_A(\mathit{Diff}_0\, \mathbf{S})\wedge\mathcal{G}(\mathbf{P}(\mathbf{S},\mathrm{ISO}^{\uparrow}(3,1))$, (9.12)

on any physical moving frame $(\boldsymbol{s},\mathbf{g}^{\mathrm{M}})$, given by a cross-section $\boldsymbol{s}$ of the Poincaré coframe bundle $P^*\mathbf{S}^{\uparrow}(\mathbf{g}^{\mathrm{M}})$ corresponding to some mean metric $\mathbf{g}^{\mathrm{M}} \in (\mathbf{S},\mathcal{g}^{\mathrm{M}})$. The $\phi$-component of this transformation can be related to the supergauge transformation

$$\mathbf{U}_{\boldsymbol{s}}(\boldsymbol{\psi}(x),\boldsymbol{\phi}(x)) = \boldsymbol{U}(a(x),R(x))\otimes\mathbf{U}(\boldsymbol{s}(x);\bar{\boldsymbol{c}}(x),\boldsymbol{b}(x),\boldsymbol{c}(x)) \ , \tag{9.13}$$

whose first factor is determined on the basis of Eqs. (7.6) and (8.1),

$$\boldsymbol{U}(a,R) = \bigoplus\nolimits_{n=0}^{\infty} U(a,R)^{\otimes n} \ , \qquad (a,R) \in \mathrm{T}^4 \wedge \mathrm{O}(3) \subset \mathrm{ISO}^{\uparrow}(3,1) \ , \tag{9.14}$$

whereas the second factor results from expressing the quantum gravitational coframe in Eq. (9.4) in the form

$$\boldsymbol{\Phi}^{\dagger}_{\bar{\mathbf{c}},\mathbf{b},\mathbf{c}} = \boldsymbol{\Psi}_{0;\boldsymbol{u}} \otimes \boldsymbol{\Psi}^{\dagger}_{0;\boldsymbol{u}} \cdot \mathbf{U}(\boldsymbol{u};\bar{\boldsymbol{c}},\boldsymbol{b},\boldsymbol{c}) \ , \tag{9.15}$$

$$\mathbf{U}(\boldsymbol{u};\bar{\boldsymbol{c}},\boldsymbol{b},\boldsymbol{c}) = \exp\Big(\theta\bar{C}(\boldsymbol{u}) + \bar{\theta}C(\boldsymbol{u}) + \theta\bar{\theta}\Big[B(\boldsymbol{u}) + \tfrac{1}{2}C(\boldsymbol{u})\times\bar{C}(\boldsymbol{u})\Big]\Big) \ , \tag{9.16}$$

$$\bar{\theta}C(\boldsymbol{u}) = \bar{\theta}^{\alpha}C_{\alpha,\mathbf{0}}(\boldsymbol{u}), \quad \theta\bar{C}(\boldsymbol{u}) = \theta^{\alpha}\bar{C}_{\alpha}(\boldsymbol{u}) \ , \quad \theta\bar{\theta}B(\boldsymbol{u}) = \sum\nolimits_{\alpha=1}^{\infty}\theta^{\alpha}\bar{\theta}^{\alpha}B_{\alpha}(\boldsymbol{u}) \ , \tag{9.17}$$

$$\theta\bar{\theta}C(\boldsymbol{u})\times\bar{C}(\boldsymbol{u}) = \sum\nolimits_{\alpha=1}^{\infty}\theta^{\alpha}\bar{\theta}^{\alpha}\Big(4 + \{C_{\alpha,\mathbf{0}}(\boldsymbol{u}),\bar{C}_{\alpha}(\boldsymbol{u})\}\Big) , \quad B_{\alpha}(\boldsymbol{u}) = B(\boldsymbol{u};\mathrm{b}^{\alpha}\mathbf{c}_{\boldsymbol{u};\alpha}) \ , \tag{9.18}$$

where $B$ denotes a bosonic gauge-fixing field, and $C$ and $\bar{C}$ are the ghost and antighost fields whose creation operators appear in Eqs. (8.12) and (8.13), respectively. Thus, if we start from a TT-gauge fixing with respect to $(\boldsymbol{s},\mathbf{g}^{\mathrm{M}})$, which corresponds to a choice of quantum gravitational coframes whose elements are the duals of the coherent states $\boldsymbol{\Phi}_{\boldsymbol{f}}$ for $\boldsymbol{f} \in \mathbf{F}_{\boldsymbol{s}(x)}$ in the TT-gauge everywhere in $\mathbf{S}$, and take advantage of Eqs. (9.1)-(9.2) to write

$$\boldsymbol{\phi}\big(\boldsymbol{s}(x)\cdot(a(x),\Lambda(x))\big) = (0,\Lambda_v^{-1}(x))\,\boldsymbol{\phi}\big(\boldsymbol{s}(x)\cdot(a(x),R(x))\big)(0,\Lambda_v(x)) \ , \quad x \in \mathbf{S} \ , \tag{9.19}$$

we can assign to the gauge transformation in Eq. (9.12) the *quantum* gravitational gauge transformation provided by the following change of quantum gravitational coframes:

$$\boldsymbol{\Phi}^{\dagger}_{\boldsymbol{f}} \ \mapsto \ \boldsymbol{\Phi}^{\dagger}_{\boldsymbol{f}}\cdot\mathbf{U}_{\boldsymbol{s}}(\boldsymbol{\psi},\boldsymbol{\phi}) = \boldsymbol{\Phi}^{\dagger}_{\boldsymbol{f}}\cdot\boldsymbol{U}(a,R)\otimes\boldsymbol{\Psi}^{\dagger}_{0;\boldsymbol{u}}\cdot\mathbf{U}(\boldsymbol{u};\bar{\boldsymbol{c}},\boldsymbol{b},\boldsymbol{c}) = \boldsymbol{\Phi}^{\dagger}_{U^{-1}(a,R)\boldsymbol{f},\bar{\mathbf{c}},\mathbf{b},\mathbf{c}} \ . \tag{9.20}$$

By carrying out the resulting the transformations for all physical moving frames $\boldsymbol{s}$ in $P^*\mathbf{S}^{\uparrow}(\mathcal{g}^{\mathrm{M}})$, all diffeomorphisms $\psi \in \mathit{Diff}_0\mathbf{S}$, and all Poincaré gauge transformations $\phi \in \mathcal{G}(\mathbf{P}(\mathbf{S},\mathrm{ISO}^{\uparrow}(3,1))$, we arrive at the action of a quantum gravitational supergroup $\mathcal{Q}\mathcal{G}(\mathbf{S},\mathcal{g}^{\mathrm{M}})$ upon all quantum gravitational coframes over $\mathbf{S}$.

## 10. The Quantum Gravitational Connection

To be able to take into account during the formulation of quantum geometrodynamic evolution the degrees of freedom contained in the Grassmann variables in Eqs. (9.15)-(9.18), we associate with each base-segment $\mathbf{S}$ the supermanifold

$$\boldsymbol{S} = \Big\{(x,\theta,\bar{\theta})\,\Big|\, x \in \mathbf{S}, \ \ \theta = (\theta^1,\theta^2,\ldots), \ \ \bar{\theta} = (\bar{\theta}^1,\bar{\theta}^2,\ldots)\Big\} \ . \tag{10.1}$$

This base supermanifold contains the *quantum* spacetime base-segment

$$\boldsymbol{S}(\mathcal{g}^{\mathrm{M}}) = \Big\{(x,\theta,\bar{\theta})\,\Big|\, x \in \mathbf{S}, \ \theta = (\theta^1_{\boldsymbol{s}(x)},\ldots), \ \bar{\theta} = (\bar{\theta}^1_{\boldsymbol{s}(x)},\ldots), \ \boldsymbol{s} \in \Sigma(P^*\mathbf{S}^{\uparrow}(\mathcal{g}^{\mathrm{M}}))\Big\} \ , \tag{10.2}$$

obtained as $\boldsymbol{s}$ ranges over the family $\Sigma(P^{*}\mathbf{S}^{\uparrow}(g^{\mathrm{M}}))$ of all cross-sections of the metrized coframe bundle $P^{*}\mathbf{S}^{\uparrow}(g^{\mathrm{M}})$. Consequently, in this supermanifold we can introduce the quantum gravitational connection

$$\boldsymbol{\nabla} = \mathbf{d} + \boldsymbol{\Gamma} + \boldsymbol{C} + \overline{\boldsymbol{C}} \quad , \qquad \mathbf{d} = \boldsymbol{d} + \boldsymbol{\delta} + \overline{\boldsymbol{\delta}} \quad , \tag{10.3}$$

which incorporates the BRST and anti-BRST operators $\delta$ and $\overline{\delta}$,

$$\boldsymbol{d} = \boldsymbol{\theta}^{j}\, \boldsymbol{\partial}_{j} = \boldsymbol{d}x^{\mu}\; \partial/\partial x^{\mu}, \qquad \boldsymbol{\delta} = \boldsymbol{d}\overline{\theta}^{\alpha}\; \partial/\partial\overline{\theta}^{\alpha}, \qquad \overline{\boldsymbol{\delta}} = \boldsymbol{d}\theta^{\alpha}\; \partial/\partial\theta^{\alpha}, \tag{10.4}$$

of the general type that appears in quantum theories for nonabelian fields.[32]

The values of the remaining terms in Eq. (10.3) are obtained by first working in the TT-gauge, and with quantum gravitational frames that do not as yet contain the Grassmann variables in Eq. (10.2), so that by an adaptation[10] of the techniques of Sec. 4 we can set

$$\boldsymbol{\nabla}^{\mathrm{TT}} = \boldsymbol{d} + \boldsymbol{\Gamma}^{\mathrm{TT}}, \quad \boldsymbol{\Gamma}^{\mathrm{TT}} = i\,\tilde{\boldsymbol{\theta}}^{j}_{\boldsymbol{s}} \boldsymbol{P}_{j;\boldsymbol{s}} + \tfrac{i}{2}\,\tilde{\boldsymbol{\omega}}^{\boldsymbol{s}}_{ab}\, \boldsymbol{M}^{ab}_{\boldsymbol{s}}, \quad j = 0,1,2,3\ , \quad a,b = 1,2,3\ , \tag{10.5}$$

where the infinitesimal generators of local spacetime translations and rotations

$$\boldsymbol{P}_{j;\boldsymbol{u}} = \int :T_{jk}[\boldsymbol{g}(\boldsymbol{u};\zeta)]:\, d\hat{\sigma}^{k}(q)\, d\Omega(v) \ , \qquad j = 0,1,2,3\ , \tag{10.6}$$

$$\boldsymbol{M}^{ij}_{\boldsymbol{u}} = \int :Q^{i}_{\boldsymbol{u}} T^{jk}[\boldsymbol{g}] - Q^{j}_{\boldsymbol{u}} T^{ik}[\boldsymbol{g}] + \boldsymbol{g}^{\dagger}\, S^{ijk}_{\boldsymbol{u}}\, \boldsymbol{g} :\, d\hat{\sigma}_{k}(q)\, d\Omega(v)\ , \tag{10.7}$$

are derivable from the following quantum gravitational stress-energy tensor:[9,10]

$$T_{ij}[\boldsymbol{g}] = \tfrac{1}{2}\, \eta_{ij}\, g_{kl,m}\, g^{kl,m} - g_{km,i}\, g^{km}{}_{,j}\ , \qquad g_{ik,j} = \partial g_{ik}/\partial q^{j}\ , \tag{10.8}$$

$$\boldsymbol{g} = g_{ij}\, \boldsymbol{\theta}^{i}(x) \otimes \boldsymbol{\theta}^{j}(x)\ , \qquad g_{ij} = g^{(+)}_{ij} + g^{(-)}_{ij}\ , \qquad g^{ij} = \eta^{ii'}\eta^{jj'} g_{i'j'}\ . \tag{10.9}$$

The transition to generic gauges can be then achieved by taking advantage of the Maurer-Cartan form $g^{-1}\boldsymbol{d}g$ of the quantum gravitation gauge group, which makes its appearance in the connection form

$${}^{g}\omega \;=\; \mathrm{Ad}_{g^{-1}}\, \omega + g^{-1}\boldsymbol{d}g\ , \qquad g = (\psi,\phi) \in \mathcal{G}(\mathbf{S}, g^{\mathrm{M}})\ , \tag{10.10}$$

for the quantum gravitational connection. Hence, in a generic Poincaré gauge $\boldsymbol{s} \in \Sigma(P^{*}\mathbf{S}^{\uparrow}(\mathbf{g}^{\mathrm{M}}))$ we have (cf. Ref. 10, Sec. 8.7):

$$\boldsymbol{\nabla}_{j} = \boldsymbol{\partial}_{j} + \boldsymbol{\Gamma}_{j}\ , \qquad \boldsymbol{\Gamma}_{j} = \mathrm{Ad}_{\mathbf{U}^{-1}(g)}\, \boldsymbol{\Gamma}^{\mathrm{TT};\boldsymbol{s}}_{j} + \mathbf{U}^{-1}(g)\lambda_{j}{}^{\mu}\, \boldsymbol{\partial}_{\mu}\mathbf{U}(g)\ , \tag{10.11}$$

$$\boldsymbol{\Gamma}_{j}(x,\theta,\overline{\theta}) = \boldsymbol{\Gamma}^{\mathrm{TT};\boldsymbol{s}(x)}_{j} + \theta D_{j}\overline{C}(x) + \overline{\theta} D_{j}C(x) + \theta\overline{\theta}\left(D_{j}B(x) + D_{j}C(x) \times \overline{C}(x)\right)\ , \tag{10.12}$$

$$D_{j} = \boldsymbol{\partial}_{j} + [\boldsymbol{\Gamma}^{\mathrm{TT};\boldsymbol{s}}_{j}, \cdot\,]\ , \qquad \boldsymbol{\partial}_{j} := \partial_{\boldsymbol{e}_{j}} = \lambda_{j}{}^{\mu}\, \boldsymbol{\partial}_{\mu}\ . \tag{10.13}$$

The infinitesimal parallel transport in directions related to the Grassmannian degrees of freedom labeled by $\theta$ and $\overline{\theta}$ is governed by the following quantum gravitational BRST and anti-BRST connection forms,

$$s = \boldsymbol{\delta} + U^{-1}(g)\, \boldsymbol{\partial}_{\overline{\theta}} U(g)\, \boldsymbol{d}\overline{\theta}\ , \qquad \boldsymbol{\partial}_{\overline{\theta}} = \partial/\partial\overline{\theta}\ , \tag{10.14}$$

$$\overline{s} = \overline{\boldsymbol{\delta}} + U^{-1}(g)\, \boldsymbol{\partial}_{\theta} U(g)\, \boldsymbol{d}\theta\ , \qquad \boldsymbol{\partial}_{\theta} = \partial/\partial\theta\ , \tag{10.15}$$

derivable from Eq. (10.10). This yields the gauge potentials

$$C_{\alpha}(x,\theta,\overline{\theta}) = C_{\alpha}(x) + \theta^{\alpha}\overline{B}_{\alpha}(x) - \tfrac{1}{2}\overline{\theta}^{\alpha}\{C_{\alpha}(x), C_{\alpha}(x)\} + \theta^{\alpha}\overline{\theta}^{\alpha}[\overline{B}_{\alpha}(x), C_{\alpha}(x)]\ , \tag{10.16}$$

$$\overline{C}_\alpha(x,\theta,\overline{\theta}) = \overline{C}_\alpha(x) + \overline{\theta}^\alpha B_\alpha(x) - \tfrac{1}{2}\theta^\alpha\{\overline{C}_\alpha(x),\overline{C}_\alpha(x)\} - \theta^\alpha\overline{\theta}^\alpha[B_\alpha(x),\overline{C}_\alpha(x)] \ , \quad (10.17)$$

in which we have set, by definition,

$$\overline{B}(x) = -B(x) - C(x)\times\overline{C}(x) \ . \quad (10.18)$$

The quantum gravitational connection described by Eqs. (10.3)-(10.18) governs the parallel transport of quantum gravitational superframes that underlies the quantum geometrodynamic evolution of the gravitational field. For this parallel transport to be *physical* within a base supermanifold $\boldsymbol{S}$ containing Grassmannian degrees of freedom related to unphysical modes of graviton polarization it has to take place in $\boldsymbol{S}$ only along curves whose tangent vectors

$$\boldsymbol{Y} = \dot{x}^\mu\boldsymbol{\partial}_\mu + \dot{\overline{\theta}}\boldsymbol{\partial}_{\overline{\theta}} + \dot{\theta}\boldsymbol{\partial}_\theta \in T_X\,\boldsymbol{S}(\mathcal{g}^{\mathrm{M}}) \ , \qquad X = (x,\theta,\overline{\theta}) \ , \quad (10.19)$$

satisfy the following subsidiary conditions:

$$s_{\boldsymbol{Y}(t)}\,\boldsymbol{\Phi}^{\boldsymbol{s}}_{\boldsymbol{f}(t)} = \left(\overline{\partial}_\alpha + C_\alpha(X(t))\right)\dot{\overline{\theta}}^\alpha(t)\,\boldsymbol{\Phi}^{\boldsymbol{s}}_{\boldsymbol{f}(t)} = \boldsymbol{0} \ , \quad (10.20)$$

$$\overline{s}_{\boldsymbol{Y}(t)}\,\boldsymbol{\Phi}^{\boldsymbol{s}}_{\boldsymbol{f}(t)} = \left(\partial_\alpha + \overline{C}_\alpha(X(t))\right)\dot{\theta}^\alpha(t)\,\boldsymbol{\Phi}^{\boldsymbol{s}}_{\boldsymbol{f}(t)} = \boldsymbol{0} \ . \quad (10.21)$$

This is a geometric version of the Kugo-Ojima[34] type of subsidiary conditions that is encountered also in the conventional treatment of quantum nonabelian gauge fields.[32] In the present quantum-geometric context it has the following physical meaning: for any fixed choice of gauge for the diffeomorphism group, reflected in a choice of mean metric $\mathbf{g}^{\mathrm{M}}$ from the equivalence class $\mathcal{g}^{\mathrm{M}}$ of diffeomorphically equivalent mean metrics, and for any fixed Poincaré gauge, corresponding to some choice $\boldsymbol{s}$ of cross-section of the Poincaré coframe bundle $P^*\mathbf{S}^{\uparrow}(\mathbf{g}^{\mathrm{M}})$, all *physical* parallel transport underlying quantum-geometric propagation in quantum spacetime has to take place along paths which are horizontal lifts of paths in base-segments $\mathbf{S}$, so that the causality prevailing in each $\mathbf{S}$ is thereby preserved.

## 11. Quantum Geometrodynamic Evolution

In order to formulate the quantum geometrodynamic evolution of a superlocal quantum gravitational field in mutual interaction with all "matter" quantum fields in existence, we modify the pattern followed in Sec. 5 to suit the present much more general situation. Thus, we first introduce the gravitational exciton-mode fibres

$$\hat{\mathbf{Z}}^{\mathrm{TT}}_{\boldsymbol{u}} = \left\{\hat{\boldsymbol{f}}\ \middle|\ \boldsymbol{f}\in\mathbf{Z}^{\mathrm{TT}}_{\boldsymbol{u}}\right\} \ , \qquad \hat{\boldsymbol{f}}(v) := \boldsymbol{f}(\boldsymbol{u};-a(x)-i\ell v) \ , \qquad v\in\boldsymbol{V}^+ , \quad (11.1)$$

whose indefinite inner product is formally given by

$$\left\langle\hat{\boldsymbol{f}}\middle|\hat{\boldsymbol{f}}'\right\rangle = Z_{\ell,0}\int_{\boldsymbol{V}^+}\eta^{ii'}\eta^{jj'}\hat{f}^*_{ij}(v)\,\hat{f}'_{i'j'}(v)\,d\Omega(v) = \langle\boldsymbol{f}|\boldsymbol{f}'\rangle \ , \quad (11.2)$$

where the renormalization constant $Z_{\ell,0}$ is infinite – so that the above inner product is *de facto* determined by a zero-mass limit.[9,10] The elements of the ensuing quantum gravitational field-mode fibres,

$$\hat{\mathcal{Z}}^{\mathrm{TT}}_{\boldsymbol{u}} = \bigoplus_{n=0}^{\infty}\hat{\mathcal{Z}}^{\mathrm{TT}}_{\boldsymbol{u};n} \ , \qquad \hat{\mathcal{Z}}^{\mathrm{TT}}_{\boldsymbol{u};n} = \hat{\mathbf{Z}}^{\mathrm{TT}}_{\boldsymbol{u}}\underset{S}{\otimes}\cdots\underset{S}{\otimes}\hat{\mathbf{Z}}^{\mathrm{TT}}_{\boldsymbol{u}} \ , \quad (11.3)$$

can be identified, as in Eq. (5.7), with those of the corresponding TT-subfibres of $\mathcal{Z}_{\boldsymbol{u}}$,:

$$\hat{\Psi}_n(v_1,\ldots,v_n) = \Psi_n(\boldsymbol{u};-a(x)-i\ell v_1,\ldots,-a(x)-i\ell v_n) \ , \quad \Psi_n \in \hat{\mathcal{Z}}^{\mathrm{TT}}_{\boldsymbol{u};n} \ . \tag{11.4}$$

An analogous procedure can be applied to ghost and antighost field-modes, thus ultimately arriving at the quantum gravitational field-mode superfibres

$$\hat{\mathcal{Z}}^{\lozenge}_{\boldsymbol{s}(x)} = \hat{\mathcal{Z}}^{\mathrm{TT}}_{\boldsymbol{s}(x)} \otimes \hat{\overline{\mathcal{B}}}_{\boldsymbol{s}(x)} \otimes \hat{\mathcal{B}}_{\boldsymbol{s}(x)} \ . \tag{11.5}$$

Within these superfibres we can implement the continuous expansions

$$\left\langle \hat{\boldsymbol{\Psi}} \middle| \hat{\boldsymbol{\Psi}}' \right\rangle = \int_{\mathbf{Z}^{\mathrm{TT}}_{\boldsymbol{s}(x)}} \boldsymbol{df}\,\boldsymbol{d\bar{f}} \int \boldsymbol{d\bar{\theta}}\,\boldsymbol{d\theta} \left\langle \hat{\boldsymbol{\Phi}}^{\boldsymbol{s}}_{\boldsymbol{f}}(x,\theta,\bar{\theta}) \middle| \hat{\overline{\boldsymbol{\Psi}}} \otimes \hat{\boldsymbol{\Psi}}' \right\rangle \ , \quad \hat{\boldsymbol{\Psi}}, \hat{\boldsymbol{\Psi}}' \in \hat{\mathcal{Z}}^{\mathrm{TT}}_{\boldsymbol{s}(x)} \otimes \hat{\mathcal{F}}_{\boldsymbol{s}(x)} \ , \tag{11.6}$$

by the methods of Berezin integration.

For any given choice of a physical moving frame $\boldsymbol{s} = \{(\boldsymbol{a}^*(x), \theta^i(x)) \,|\, x \in \mathbf{S}\}$, we define a *quantum gravitational superfield* at each base superlocation $X$ as follows:

$$\mathbf{g}^{\mathrm{Q}}(X,v) = \mathrm{g}^{\mathrm{Q}}_{ij}(X,v)\, \boldsymbol{\theta}^i(x) \otimes \boldsymbol{\theta}^j(x) \ , \qquad X = (x,\theta,\bar{\theta}) \in \boldsymbol{S}(\mathcal{g}^{\mathrm{M}}) \ , \tag{11.7}$$

$$\mathrm{g}^{\mathrm{Q}}_{ij}(X,v) = \int_{\mathbf{Z}^{\mathrm{TT}}_{\boldsymbol{s}(x)}} \boldsymbol{df}\,\boldsymbol{d\bar{f}} \int_{\mathbf{Z}^{\mathrm{TT}}_{\boldsymbol{s}(x)}} \boldsymbol{df'}\,\boldsymbol{d\bar{f}'} \left| \hat{\boldsymbol{\Phi}}^{\boldsymbol{s}}_{\boldsymbol{f}}(X) \right\rangle \mathrm{g}^{\mathrm{Q}}_{ij\boldsymbol{f}\boldsymbol{f}'}(x,v) \left\langle \hat{\boldsymbol{\Phi}}^{\boldsymbol{s}}_{\boldsymbol{f}'}(X) \right| \ , \tag{11.8}$$

$$\mathrm{g}^{\mathrm{Q}}_{ij\boldsymbol{f}\boldsymbol{f}'}(x,v) = \left\langle \hat{\boldsymbol{\Phi}}_{\boldsymbol{f}} \middle| \left(\eta_{ij} + g_{ij}(\boldsymbol{s}(x);-i\ell v)\right) \hat{\boldsymbol{\Phi}}_{\boldsymbol{f}'} \right\rangle \ , \quad \hat{\boldsymbol{\Phi}}_{\boldsymbol{f}}, \hat{\boldsymbol{\Phi}}_{\boldsymbol{f}'} \in \hat{\mathcal{Z}}^{\mathrm{TT}}_{\boldsymbol{s}(x)} \ . \tag{11.9}$$

We can then easily establish that the quantum gravitational fluctuations around the mean quantum gravitational field values

$$\mathbf{g}^{\mathrm{M}}(x) = \eta_{ij}\, \boldsymbol{\theta}^i(x) \otimes \boldsymbol{\theta}^j(x) \ , \qquad x \in \mathbf{S} \ , \tag{11.10}$$

are provided by the quantum gravitational *radiation* superfield with components

$$\mathrm{g}_{ij}(X,v) = \mathbf{U}^{\dagger}_{\boldsymbol{s}}(X)\, \mathrm{g}_{ij}(x,v)\, \mathbf{U}_{\boldsymbol{s}}(X) \ , \qquad X = (x,\theta,\bar{\theta}) \in \boldsymbol{S}(\mathcal{g}^{\mathrm{M}}) \ , \tag{11.11}$$

$$\mathrm{g}_{ij}(x,v) = \int_{\mathbf{Z}^{\mathrm{TT}}_{\boldsymbol{s}(x)}} \boldsymbol{df}\,\boldsymbol{d\bar{f}} \int_{\mathbf{Z}^{\mathrm{TT}}_{\boldsymbol{s}(x)}} \boldsymbol{df'}\,\boldsymbol{d\bar{f}'} \left| \hat{\boldsymbol{\Phi}}_{\boldsymbol{f}} \right\rangle \left( \hat{f}^{*}_{ij}(v) + \hat{f}'_{ij}(v) \right) \left\langle \hat{\boldsymbol{\Phi}}_{\boldsymbol{f}} \middle| \hat{\boldsymbol{\Phi}}_{\boldsymbol{f}'} \right\rangle \left\langle \hat{\boldsymbol{\Phi}}_{\boldsymbol{f}'} \right| \ . \tag{11.12}$$

To describe the mutual interaction of "matter" fields with the quantum gravitational superfield in Eq. (11.7), these fields have to be defined at all superlocations $X \in \boldsymbol{S}(\mathcal{g}^{\mathrm{M}})$. This can be achieved by "lifting," for example, the coherent field-mode sections in Eq. (5.10) from the base-locations $x$ to the base-superlocations $X$,

$$\hat{\boldsymbol{\Phi}}^{\boldsymbol{s}}_{\varphi(x)} \mapsto \hat{\boldsymbol{\Phi}}^{\boldsymbol{s}}_{\varphi(X)} \in \hat{\mathcal{F}}^{\mathrm{mat}}_X \ , \qquad \varphi(x) \in \mathbf{F}^{\mathrm{mat}}_x \ , \quad \varphi(X) = \mathbf{U}^{-1}_{\boldsymbol{s}}(X)\, \varphi(x) \in \mathbf{F}^{\mathrm{mat}}_X \ , \tag{11.13}$$

when executing a quantum gravitational gauge transition involving Lorentz boosts. More general situations can be handled along similar lines, thus arriving at well-defined proper energy-momentum operators for "matter" fields,

$$\hat{\boldsymbol{P}}^{\mathrm{mat}}_{j;\boldsymbol{s}}(X) = n^k(x) \int_{\boldsymbol{V}^+_x} :T^{\mathrm{mat}}_{jk}[\boldsymbol{\varphi}^{\mathrm{mat}}(X,\boldsymbol{v})]: \, d\Omega(\boldsymbol{v}) \ , \qquad j = 0,1,2,3 \ . \tag{11.14}$$

These operators are obtained from the stress-energy tensors in the semi-classical regime by replacing in the latter Poincaré-covariant derivatives with corresponding quantum gravitationally covariant derivatives, such as

$$\begin{aligned} \boldsymbol{\nabla}^{\mathrm{mat}}_j &= \boldsymbol{\partial}_j - i\tilde{\boldsymbol{\theta}}^j_{\boldsymbol{s}} \boldsymbol{P}^{\mathrm{mat}}_{j;\boldsymbol{s}} + \tfrac{i}{2} \tilde{\boldsymbol{\omega}}^{\boldsymbol{s}}_{ab} \boldsymbol{M}^{ab}_{\mathrm{mat};\boldsymbol{s}} \\ &+ \theta D_j \overline{C}(x) + \bar{\theta} D_j C(x) + \theta\bar{\theta}\left( D_j B(x) + D_j C(x) \times \overline{C}(x) \right) \ . \end{aligned} \tag{11.15}$$

The resulting expressions contribute to the real-valued part of the total mean (local)

proper energy density and total mean (local) proper 3-momentum density:

$$\bar{\boldsymbol{\rho}}^{\text{tot}}(x) = \left\langle \boldsymbol{\Psi}_X^{\text{tot}} \middle| \boldsymbol{\rho}^{\text{tot}}(X)\,\boldsymbol{\Psi}_X^{\text{tot}} \right\rangle^{\text{body}}, \qquad \bar{\boldsymbol{j}}_a^{\text{tot}}(x) = \left\langle \boldsymbol{\Psi}_X^{\text{tot}} \middle| \boldsymbol{j}_a^{\text{tot}}(X)\,\boldsymbol{\Psi}_X^{\text{tot}} \right\rangle^{\text{body}}. \tag{11.16}$$

In turn, the above values of these densities along each given inflow hypersurface of a base-segment, together with those of the mean Riemannian 3-metric tensor $\gamma^{\mathrm{M}}$ induced by $\mathbf{g}^{\mathrm{M}}$ and of the mean extrinsic curvature tensor $\boldsymbol{K}$, dictate the creation of the infinitesimal QGR gauge orbits $(\mathbf{S}, \mathcal{g}^{\mathrm{M}})$. Mathematically, each gauge orbit is obtained by solving[24] the CGR Cauchy problem, with the four Einstein equations

$$R(\boldsymbol{\gamma}^{\mathrm{M}}) - K_{ab}K^{ab} + \left(K^a{}_a\right)^2 = 16\pi\bar{\boldsymbol{\rho}}^{\text{tot}}, \qquad K^a{}_b = \gamma^{\mathrm{M}}_{bc}K^{ac}, \tag{11.17}$$

$$\mathrm{D}_b K^b{}_a - \mathrm{D}_a K^b{}_b = 8\pi\bar{\boldsymbol{j}}_a^{\text{tot}}, \qquad a = 1,2,3, \tag{11.18}$$

used as constraints on $\boldsymbol{K}$ for given $\gamma^{\mathrm{M}}$, where $D_a$ denotes the $\boldsymbol{e}_a$-component (with respect to the $\gamma^{\mathrm{M}}$-orthonormal spatial frames $\{\boldsymbol{e}_a(x)|\, a = 1,2,3\}$ dual to $\{\theta^a(x)|\, a = 1,2,3\}$) of the operator for the covariant differentiation determined by the Levi-Civita connection to which each 3-metric $\gamma^{\mathrm{M}}$ gives rise on the inflow hypersurface of $\mathbf{S}$.

Thus, the framework for geometric quantum gravity postulates no background metrics, and, in fact, it does not assume that a classical spacetime manifold of any kind is *a priori* given. However, it does take for granted that, in accordance with observational data, mean metrics constituting QGR gauge orbits perpetually emerge during the process of quantum-geometric evolution as a result of the metrization of classical frame and coframe bundles.

The self-interactions of the quantum gravitational radiation field in Eq. (11.11) are taken to contribute, following a study by Deser,[35] to the total operator-valued densities in Eq. (11.16) via the self-interaction stress-energy tensor

$$T_{ij}^{\text{s.int}}[\mathbf{g}] = (1/16\pi)\left(\hat{\Gamma}^l{}_{ij}\hat{\Gamma}^k{}_{kl} - \hat{\Gamma}^l{}_{ki}\hat{\Gamma}^k{}_{jl}\right), \qquad \hat{\Gamma}^i{}_{jk} = \tfrac{1}{2}\eta^{il}\left(\mathrm{g}_{lj,k} + \mathrm{g}_{lk,j} - \mathrm{g}_{jk,l}\right). \tag{11.19}$$

The above stress-energy tensor also contributes to the total stress-energy tensor for the energy-momentum of quantum gravitational radiation

$$\hat{\boldsymbol{P}}_{j;\boldsymbol{s}}^{\text{rad}}(X) = n^k(x)\int_{\boldsymbol{V}_x^+} :T_{jk}^{\text{rad}}[\mathbf{g}(X,\boldsymbol{v})]:\, d\Omega(\boldsymbol{v}), \qquad j = 0,1,2,3, \tag{11.20}$$

a term which has to be added to the one in (10.8), so that:

$$T_{ij}^{\text{rad}}[\mathbf{g}] = T_{ij}[\mathbf{g}] + T_{ij}^{\text{s.int}}[\mathbf{g}]. \tag{11.21}$$

The mean values in Eq. (11.16) resulting from Eqs. (11.14) and (11.20) along each inflow hypersurface $\Sigma_{t_{n-1}}$ determine (cf. Ref. 24, Sec. 1.2.5), via the constraint equations (11.17)-(11.18) and the remaining six geometrodynamic Einstein equations for $\mathbf{g}^{\mathrm{M}}$, the QGR gauge orbit $(\mathbf{S}_n, \mathcal{g}^{\mathrm{M}})$, including the outflow hypersurface $\Sigma_{t_n}$ of $\mathbf{S}_n$ at the proper time separation $t_n - t_{n-1} = (t'' - t')/N = \varepsilon$. Hence, the quantum spacetime segment $\boldsymbol{S}_n(\mathcal{g}^{\mathrm{M}})$ is thereby determined via Eqs. (10.1) and (10.2).

With all the remaining ingredients of the quantum-geometric framework for gravity now available, the formulation of the quantum-geometric propagator

$$K(f(X''); f(X')) = \lim_{\varepsilon \to +0} \prod_{n=N}^{1}{}' \int \mathcal{D}f_n \left\langle \hat{\boldsymbol{\Phi}}_{f_n}^{\boldsymbol{s}}(X_n) \middle| e^{-i\varepsilon\int d\sigma(x_n)\,\boldsymbol{\rho}^{\text{tot}}(X_n)}\; \hat{\boldsymbol{\Phi}}^{\boldsymbol{s}}(X_n; f_{n-1}) \right\rangle, \tag{11.22}$$

$$\mathcal{D}f_n = \prod_{x_n \in \Sigma_{t_n}} \mathcal{D}[\mathrm{g}_n(x_n)]\, \mathcal{D}[\bar{c}_n(x_n)]\, \mathcal{D}[c_n(x_n)]\mathcal{D}[\varphi_n^{\text{mat}}(X_n)]. \tag{11.23}$$

for gravitational and "matter" superfields, whose exciton modes are collectively denoted by $f$, can be obtained by suitable adaptation of the type of iterative procedures considered in Sec. 5. The Grassmannian values assumed by the ghost and antighost exciton modes above each base location $x_n \in \Sigma_{t_n}$ are those corresponding to various superlocations $X_n \in \Sigma_{t_n}$. Thus, the quantum geometrodynamic propagation proceeds, under the subsidiary conditions in Eqs. (10.20) and (10.21), along those paths $\gamma^{\uparrow}(X_{n-1}, X_n)$ in the base-supersegments $\boldsymbol{S}_n(g^{\mathrm{M}})$ which are the lifts of the geodesic arcs $\gamma(x_{n-1}, x_n)$ connecting each point $x_{n-1}$ on the inflow hypersurface $\Sigma_{t_{n-1}}$ with the points $x_n$ on the outflow hypersurface $\Sigma_{t_n}$ of each base-segment $\mathbf{S}_n$.

## 12. Conclusion

We can now in the position to compare the framework for geometric quantum gravity, outlined in the preceding five sections, with that of the covariant and the canonical frameworks[36] for quantum gravity.

The geometric formulation shares with the covariant one the use of graviton states representing quantum gravitational radiation. On the other hand, the covariant formulation postulates a globally existing background metric – for which, on pragmatic grounds, the usual choice is that of Minkowski metric.[36] In addition to the loss of diffeomorphism as well as Poincaré gauge invariance, such a postulate precludes the *mutual* interaction between matter and geometry, which is physically essential to general relativity.

Consequently, in the geometric framework the *mean* metrics are those *created* by that mutual interaction during the quantum geometrodynamic evolution. This allows the indeterministic quantum phenomena to affect their values locally. Hence, the resulting quantum spacetime has to be viewed as being in a constant process of creation, rather than as a conceptually static object, determined in its minutest details at the instant of its creation.

As in the canonical approach to quantum gravity,[36,37] this *quantum* geometrodynamic evolution requires reference 3-surfaces. Quantum base geometries are then created by the metrization of the general affine frame bundles over those surfaces. However, whereas in the superspace approach to quantum gravity the fluctuations of 3-metrics are envisaged to take place globally over those reference 3-surfaces, in the geometric approach such metric fluctuations take place locally, i.e., within the fibres above their base spacetime locations.

These dissimilarities become more pronounced in relation to recent work on canonical quantum gravity,[29] in which the concept of connection takes precedence over that of metric; whereas in the quantum geometric framework the concept of metric retains a central role, which remains on par with that which it plays in classical general relativity.

In general, during the entire period of development[4,9,10] of the quantum geometry framework, *physical* principles were deemed primary. Hence, suitable mathematical techniques were searched for, and, if not found in existing literature, created in order to fit those physical principles – rather than vice versa. The central physical and mathematical ideas which thereby emerged are those of *quantum (super)frame bundles* and of *quantum geometric evolution*. Quantum frame bundles represent systems of covariance[9,10] that are generalizations of the systems of imprimitivity[4,6] to which complete sets of observables give rise; whereas quantum geometric evolution represents a physically grounded geometric extrapolation of the path integral method to the general relativistic regime. Thus, the entire quantum geometric framework for quantum field theory in curved spacetime and quantum gravity rests on physically well founded extrapolations of orthodox quantum theory.

## References

1. E. Prugovečki, *Ann. Phys. (N.Y.)* **110**, 102 (1978).
2. E. Prugovečki, *J. Math. Phys.* **19**, 2260 (1978).
3. E. Prugovečki, *Phys. Rev. D* **18**, 3655 (1978).
4. E. Prugovečki, *Stochastic Quantum Mechanics and Quantum Spacetime* (Reidel, Dordrecht, 1984; reprinted, with corrections, 1986).
5. M. Born, *Rev. Mod. Phys.* **21**, 463 (1949).
6. E. P. Wigner, in *Quantum Theory and Measurement*, J. A. Wheeler and W. H. Zurek, eds. (Princeton University Press, Princeton, 1983), pp. 260-314.
7. E. Prugovečki, *Class. Quantum Grav.* **4**, 1659 (1987).
8. E. Prugovečki, *Nuovo Cimento A* **97**, 597, 837 (1987).
9. E. Prugovečki, *Quantum Geometry* (Kluwer, Dordrecht, 1992).
10. E. Prugovečki, *Principles of Quantum General Relativity* (World Scientific, Singapore, 1995).
11. S. T. Ali and G. G. Emch, *J. Math. Phys.* **27**, 2936 (1986).
12. B. Kostant, *Quantization and Unitary Representations* (Springer Lecture Notes in Mathematics, vol. 170, New York, 1970).
13. J.-M. Souriau, *Structure des systèmes dynamiques* (Dunod, Paris, 1970).
14. S. T. Ali and E. Prugovečki, *Acta Appl. Math.* **6**, 1, 19, 47 (1986).
15. A. M. Perelomov, *Generalized Coherent States and Their Applications*, (Springer, Berlin, 1986).
16. F. A. Berezin, *Commun. Math. Phys.* **40**, 153 (1975).
17. E. Prugovečki, *Class. Quantum Grav.* **11**, 1981 (1994).
18. E. Prugovečki, *Quantum Mechanics in Hilbert Space*, 2nd edition (Academic Press, New York, 1981).
19. E. Prugovečki, *Nuovo Cimento A* **61**, 85 (1981).
20. B. S. DeWitt, in *Gravitation: An Introduction to Current Research*, L. Witten, ed. (Wiley, New York, 1962).
21. H.-H. von Borzeszkowski and H.-J. Treder, *The Meaning of Quantum Gravity* (Kluwer, Dordrecht, 1988).
22. W. Drechsler, *Fortschr. Phys.* **23**, 449 (1984).
23. N. N. Bogolubov, A. A. Logunov and I. T. Todorov, *Introduction to Axiomatic Quantum Field Theory* (Benjamin, Reading, Mass., 1975).
24. Y. Choquet-Bruhat and J. W. York, in *General Relativity and Gravitation*, vol. 1, A. Held, ed. (Plenum, New York, 1980).
25. F. A. Berezin, *The Method of Second Quantization* (Academic Press, New York, 1966).
26. N. D. Birrell and P. C. W. Davies, *Quantum Fields in Curved Space* (Cambridge University Press, Cambridge, 1986).
27. R. P. Feynman, *Rev. Mod. Phys.* **20**, 367 (1948).
28. R. P. Feynman, *Phys. Rev.* **76**, 749, 769 (1949).
29. A. Ashtekar and J. Stachel, eds., *Conceptual Problems in Quantum Gravity* (Birkhäuser, Boston, 1991).
30. L. de Broglie, in *Perspectives in Quantum Theory*, W. Yougrau and A. van der Merwe, eds. (Dover, New York, 1979).
31. J. Bognár, *Indefinite Inner Product Spaces* (Springer, Berlin, 1974).
32. L. Baulieu, *Phys. Reports* **129**, 1 (1985).
33. E. Cartan, *La Méthode du Repère Mobile, la Théorie des Groupes Continus et les Espaces Généralisées* (Act. Sci. et Ind., #194, Hermann, 1935).
34. T. Kugo and I. Ojima, *Prog. Theor. Phys. Suppl.* **66**, 1 (1979).
35. S. Deser, *Gen. Rel. Grav.* **1**, 9 (1970).
36. C. J. Isham, in *Quantum Gravity: An Oxford Symposium*, C.J. Isham, R. Penrose and D.W. Sciama, eds. (Clarendon Press, Oxford, 1975).

37. P. G. Bergmann and A. Komar, in *General Relativity and Gravitation*, vol. 1, A. Held, ed. (Plenum, New York, 1980).
38. E. Prugovečki, "On quantum-geometric connections and propagators in curved space-time" (University of Toronto preprint – to be published).